\documentclass[showpacs]{revtex4}
\usepackage{graphicx}
\usepackage{dcolumn}
\usepackage{amsmath}
\usepackage[latin1]{inputenc}
\usepackage{graphicx, psfrag}
\usepackage{amssymb}
\usepackage[colorlinks=true, citecolor=blue, urlcolor = blue, linkcolor= red, bookmarks=true]{hyperref}
\usepackage{float}
\usepackage{amsmath}
\usepackage{amsfonts}
\usepackage{dcolumn}
\usepackage{hyperref}
\usepackage{subfigure}
\usepackage{pgfplots}
\usepackage{epstopdf}
\usepackage{booktabs}

\title{\boldmath Rotating black holes in $4D$ Einstein-Gauss-Bonnet gravity and its shadow}

\makeatletter
\def\btt#1{\texttt{\@backslashchar#1}}
\DeclareRobustCommand\bblash{\btt{\@backslashchar}} \makeatother

\begin{document}
	
	\title[]{Rotating black holes in $4D$ Einstein-Gauss-Bonnet gravity and its shadow}
	\author{Rahul Kumar$^{a}$}\email{rahul.phy3@gmail.com}
	\author{Sushant~G.~Ghosh$^{a,\;b}$} \email{sghosh2@jmi.ac.in, sgghosh@gmail.com}
	
	\affiliation{$^{a}$ Centre for Theoretical Physics, Jamia Millia
		Islamia, New Delhi 110025, India}
	\affiliation{$^{b}$ Astrophysics and Cosmology
		Research Unit, School of Mathematics, Statistics and Computer Science, University of
		KwaZulu-Natal, Private Bag 54001, Durban 4000, South Africa}

	\date{\today}
\begin{abstract}
Recently there has been a surge of interest in regularizing, a $ D \to 4 $ limit of, the Einstein-Gauss-Bonnet (EGB) gravity, and the resulting regularized $4D$ EGB gravity has  nontrivial  dynamics.  The theory admits spherically symmetric black holes generalizing the  Schwarzschild black holes. We consider the rotating  black hole in regularized $4D$ EGB gravity and discuss their horizon properties and shadow cast. The effects of the GB coupling parameter on the shape and size of shadows are investigated in the context of recent M87* observations from the EHT. Interestingly, for a given spin parameter, the apparent size of the shadow decreases and gets more distorted due to the GB coupling parameter. We find that within the finite parameter space, e.g. for $a=0.1M$, $\alpha\leq 0.00394M^2$, and within the current observational uncertainties, the rotating black holes of the  $4D$ EGB gravity are consistent with the inferred features of M87* black hole shadow.
\end{abstract}

\maketitle
\section{Introduction}
The uniqueness of the Einstein's field equations is based on the Lovelock's theorem \cite{Lovelock:1972vz}, which ensures that  for a theory of gravity with i) four-dimensional spacetime, ii) metricity, iii) diffeomorphism invariance, and iv) second-order equations of motion, Einstein's tensor along with a cosmological constant term is the only divergence-free symmetric rank-2 tensor constructed solely from the metric tensor $g_{\mu\nu}$ and its derivatives up to second differential order. However, in higher-dimensions ($HD$) spacetimes $D>4$, the Einstein-Hilbert action is not unique, and one particularly interesting example in the $HD$ is the Einstein-Gauss-Bonnet (EGB) gravity, which is motivated by the heterotic string theory \cite{Lanczos:1938sf,Lovelock:1971yv}. Lanczos \cite{Lanczos:1938sf} and Lovelock \cite{Lovelock:1971yv}, in their pioneering works, showed that the gravitational action of  EGB gravity admits quadratic corrections constructed from the curvature tensors invariants  with action given by 
\begin{equation}
S_{\text{EGB}}
=\frac{1}{16\pi G_D} \int\! d^{D\!}x \, \sqrt{-g}(\mathcal{L}_{\text{EH}}+ \alpha \, \mathcal{L}_{\text{GB}}),\label{action}
\end{equation}
with
\begin{equation}
\mathcal{L}_{\text{EH}}=R,\;\;\;\; \mathcal{L}_{\text{GB}}=R^{\mu\nu\rho\sigma} R_{\mu\nu\rho\sigma}- 4 R^{\mu\nu}R_{\mu\nu}+ R^2,
\end{equation}
where $g$ is the determinant of $g_{\mu\nu}$, and $\alpha$ is the Gauss-Bonnet (GB) coupling constant identified as the inverse string tension and positive-definite. Varying the action (\ref{action}) with metric tensor $g_{\mu\nu}$, yields the equations of gravitational field as follows \cite{Lanczos:1938sf}
\begin{equation}
G_{\mu\nu}+\alpha H_{\mu\nu}=T_{\mu\nu},\label{FieldEq}
\end{equation}
where 
\begin{eqnarray}
G_{\mu\nu}&=&R_{\mu\nu}-\frac{1}{2}R g_{\mu\nu},\nonumber\\
H_{\mu\nu}&=&2\Bigr( R R_{\mu\nu}-2R_{\mu\sigma} {R}{^\sigma}_{\nu} -2 R_{\mu\sigma\nu\rho}{R}^{\sigma\rho} - R_{\mu\sigma\rho\delta}{R}^{\sigma\rho\delta}{_\nu}\Bigl)-\frac{1}{2}\mathcal{L}_{\text{GB}}g_{\mu\nu},
\end{eqnarray}
and $T_{\mu\nu}$ is the energy-momentum tensor for the matter fields. The EGB gravity has been widely studied, because it can be obtained in the low energy limit of string theory  \cite{Zwiebach:1985uq} and also leads to the ghost-free nontrivial gravitational self-interactions \cite{Nojiri:2018ouv}. In the EGB theories, one can explore several conceptual issues of gravity in a much broader setup than in general relativity and they are also shown to be free from instabilities when expanding about flat spacetime \cite{Boulware:1985wk, Wiltshire:1985us}.  The spherically symmetric static black hole  solution for  the EGB theory was  first  obtained by Boulware and Deser \cite{Boulware:1985wk},  later several interesting black hole solutions are obtained \cite{egb2,ghosh,egb} for various sources including the regular ones \cite{Ghosh1:2018bxg}.

For a $ D $-dimensional spacetime with $ D < 5 $, the GB Lagrangian $\mathcal{L}_{\text{GB}}$ is a total derivative and does not contribute to the gravitational dynamics. However, Tomozawa \cite{Tomozawa:2011gp} showed that, at quantum level, a consistent regularization procedure leads to the non-trivial contribution of the GB term to the $4D$ static spherically symmetric black hole spacetime, as follows 
\begin{equation}
ds^2 = -f(r)dt^2 + \frac{1}{f(r)}dr^2 +r^2(d\theta^2+\sin^2\theta\, d\phi^2), \label{NR}
\end{equation}
with
\begin{equation}
f(r)= 1 + \frac{r^2}{32 \pi \alpha G}
\Biggl[ 1\pm  \sqrt{ 1+ \frac{128 \pi\alpha  M}{r^3} } \Biggr] \, .\label{lapse}
\end{equation}
Here, $M$ is the black hole mass and $G$ is the Newton's gravitational constant which is hereafter set to be unity. Later, Cognolo et al. \cite{Cognola:2013fva} formalized  regularization procedure for EGB gravity theory by making a "entropic" dimensional reduction to $D\to 4$ within the classical Lagrangian formulation. 

However, recently, Glavan and Lin \cite{Glavan:2019inb} proposed that re-scaling the GB coupling parameter as $\alpha\to \alpha/(D-4)$ and defining the $4D$ theory as the limit of $D\to 4$ at the level of field's equation, the GB term makes a nontrivial contribution to dynamics. It has also been shown that the theory contains the degrees of freedom only of massless graviton as in GR and thus free from the ghosts \cite{Glavan:2019inb}.  Hence, the  $4D$ EGB gravity theory already attracted much attentions and being extensively studied \cite{Konoplya:2020bxa,Guo:2020zmf,Fernandes:2020rpa,Wei:2020ght,Doneva:2020ped,Ghosh:2020vpc,Singh:2020mty,Kumar:2020sag,Islam:2020xmy,Ghosh:2020syx}. Moreover, the Glavan and Lin proposal \cite{Glavan:2019inb} has been extended to four-dimensional regularization of Lovelock-Lanczos gravity up to an arbitrary curvature order \cite{Casalino:2020kbt,Konoplya:2020qqh}. 
For the spherically symmetric black hole solution (\ref{lapse}), we consider $\alpha>0$, because for the negative values of $\alpha$ eq.~(\ref{NR}) is not a valid solution at small distances \cite{Glavan:2019inb,Cognola:2013fva}. At large distances eq.~(\ref{lapse}) has two distinct branches
\begin{equation}
f(r)_-=1-\frac{2M}{r},\;\;\;
f(r)_+=1+\frac{2M}{r}+\frac{r^2}{16\pi\alpha },
\end{equation}
however, we will be focusing only on the negative branch that asymptotically goes over to the Schwarzschild black hole with correct mass sign. Whereas at small distances, the metric function (\ref{lapse}) for the negative branch reduces to
\begin{equation}
f(r)= 1 + \frac{r^2}{32 \pi \alpha}-\sqrt{\frac{M}{8\pi\alpha}}r^{1/2},
\end{equation}
which infers that the gravitational force is repulsive at small distances thereby, unlike in general relativity, an infalling particle never reaches the singularity \cite{Glavan:2019inb}.

However, several  questions are raised \cite{Ai:2020peo,Hennigar:2020lsl,Shu:2020cjw,Gurses:2020ofy,Mahapatra:2020rds,Fernandes:2020nbq} on the regularization procedure in \cite{Glavan:2019inb,Cognola:2013fva} and  also proposed  alternate ways of  regularization for $D\to 4$ \cite{Lu:2020iav,Kobayashi:2020wqy,Hennigar:2020lsl}. In particular, using the Kaluza-Klein dimensional reduction procedure \cite{Mann}, the $D\to 4$ limit of regularized EGB gravity is effectively described as a particular subclass of scalar-tensor theories within the Horndeski family \cite{Kobayashi:2020wqy,Lu:2020iav}. Interestingly, the metric (\ref{NR}) with (\ref{lapse}) emerges as a  solution of the $4D$ regularized theories \cite{Lu:2020iav,Casalino:2020kbt,Hennigar:2020lsl}.  Furthermore, the semi-classical Einstein equations with conformal anomaly proportional to Euler density $\mathcal{L}_{\text{GB}}$ \cite{Cai:2009ua} and the $4D$ non-relativistic Horava-Lifshitz theory of gravity \cite{Kehagias:2009is} also admit the identical black hole solution as in eq.~(\ref{NR}) with (\ref{lapse}).

With this motivation, it is natural and appropriate to consider  the rotating counterpart of static spherically symmetric black hole metric (\ref{NR}), namely, the rotating $4D$ EGB black hole, and discuss the horizons geometry and shadow cast by rotating black holes. Null geodesics equations of motions are obtained in the first-order differential form and the analytical expressions for the photon region are determined. Effects of the GB coupling parameter on the black hole shadow morphology is investigated, and it is found that rotating black holes in the regularized EGB gravity cast smaller and much distorted shadows than those for the Kerr black holes. Shadow observables $A$ and $D$ are used to characterize the size and shape of the shadows, and in turn to uniquely determine the black hole parameters. The M87* black hole shadow results, inferred from the recent Event Horizon Telescope (EHT) collaborations observations, are further used to constrain the GB coupling parameter.

This paper is organized as follows. In the section~\ref{sec2}, we derive the rotating EGB black holes and in turn discuss their horizon structure. Photons geodesics equations of motion and effects of the GB coupling parameter on the black hole shadow are subjects of section~\ref{sec3}.  Finally, we summarize our main findings in section~\ref{sec4}.\\

\section{Rotating black holes}\label{sec2}
The Newman$-$Janis algorithm has been widely used to construct rotating black hole solutions from their non-rotating counterparts \cite{Newman:1965tw}. While this algorithm was developed within general relativity, it has been more recently applied to non-rotating solutions in modified gravity theories, namely, some references \cite{Johannsen:2011dh,Jusufi:2019caq,Bambi:2013ufa,Ghosh:2014hea,Moffat:2014aja}. It is true that exercising the Newman$-$Janis algorithm to an arbitrary non-general relativity spherically symmetric solution introduces pathologies in the resulting axially-symmetric metric \cite{Hansen:2013owa}. More precisely in the general relativity case a source, if it exists, is the same for both a non-rotating black hole and its rotating counterpart, e.g., vacuum for both Schwarzschild and Kerr black holes, and charge for Reissner$-$Nordstrom and Kerr$-$Newman black holes. But, in cases of the  modified gravity rotating black hole counterparts, obtained using Newman$-$Janis algorithm, in addition to original sources likely have additional sources.  We derived the rotating EGB black hole metric, using the Newman$-$Janis algorithm modified by Azreg-A\"inou's non-complexification procedure\cite{Azreg-Ainou:2014pra,Azreg-Ainou:2014aqa}, which has been successfully applied for generating imperfect fluid rotating solutions in the Boyer$-$Lindquist coordinates from spherically symmetric static solutions, and can also generate generic rotating regular black hole solutions. The resulting stationary and axially symmetric (rotating) EGB black hole metric, in Boyer-Lindquist coordinates, reads
\begin{eqnarray}
ds^2 &=&  -\left(\frac{\Delta - a^2 \sin^2 \theta}{\Sigma}\right) dt^2 + \frac{ \Sigma}{\Delta }  \, dr^2  -  2 a \sin^2 \theta \left(1 - \frac{\Delta - a^2 \sin^2 \theta}{\Sigma} \right) dt \, d \phi \nonumber \\
&& + \Sigma \, d \theta^2 +  \, \sin ^2 \theta  \left[ \Sigma + a^2 \sin^2 \theta \left(2 - \frac{\Delta - a^2 \sin^2\theta}{\Sigma}\right)   \right]    d \phi^2,
\label{rotbhr}
\end{eqnarray}
with 
\begin{equation}
\Delta=r^2+a^2+\frac{r^4}{32\pi\alpha }\left[1- \sqrt{ 1+ \frac{128 \pi\alpha  M}{r^3} }\right] , \quad \Sigma=r^2+a^2\cos^2\theta,
\end{equation} 
and $a$ is the spin parameter. It may be true for our solution (\ref{rotbhr}) that it may not satisfy field equations and it is valid also for other rotating solution generated in modified gravity \cite{Johannsen:2011dh,Jusufi:2019caq,Bambi:2013ufa,Ghosh:2014hea,Moffat:2014aja}, and they likely to generate extra stresses. Therefore, we regard our rotating metric eq.~(\ref{rotbhr}) as a regularized EGB gravity black hole metric of an appropriately chosen set of field equations which are unknown but different from the EGB equations (\ref{FieldEq}).  Further, in the limit $ a = 0 $, the metric (\ref{rotbhr}) reduces to the spherically symmetric metric (\ref{NR}) \cite{Glavan:2019inb}. Nevertheless, the rotating black hole metric (\ref{rotbhr}) also corresponds to the varieties of gravity theories which admit the static black hole metric identical to eq.~(\ref{NR}) \cite{Tomozawa:2011gp,Glavan:2019inb,Cai:2009ua,Cognola:2013fva,Casalino:2020kbt,Kehagias:2009is,Konoplya:2020qqh,Hennigar:2020lsl}. Moreover, we shall demonstrate that our metric (\ref{rotbhr}) for the $4D$ regularized EGB gravity can describe rotating  black holes up to the maximum value of the spin $ a $, has rich structure, and the rotating black holes are consistent with the inferred features of M87* black hole shadow results of the EHT. The metric eq.~(\ref{rotbhr}), in the limit $\alpha\to 0$ or large $r$,  goes over to the Kerr black holes \cite{Kerr:1963ud}  and also to the the spherical symmetric solution (\ref{NR}). The rotating black hole (\ref{rotbhr}), like the Kerr black hole, possesses two linearly independent Killing vectors $\eta_{(t)}^{\mu}$ and $\eta_{(\phi)}^{\mu}$, respectively, associated with the isometries along the time translation and rotational invariance. 

\begin{figure*}
	\begin{center}
	\includegraphics[scale=0.82]{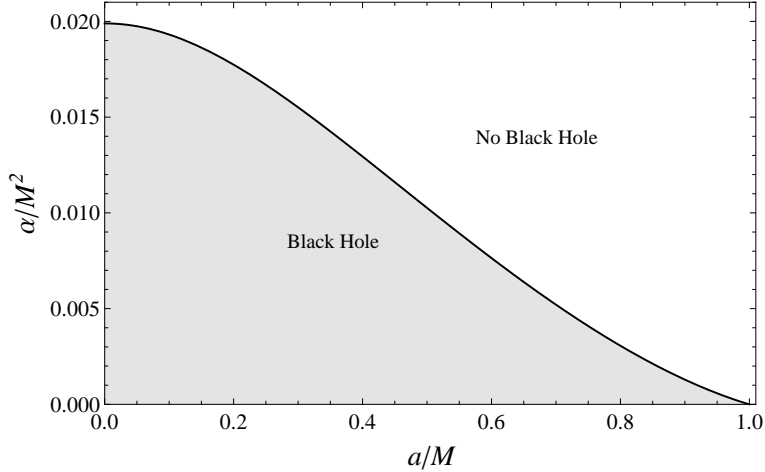}
	\end{center}
\caption{The parameter space $(a,\alpha)$ for the existence of the black hole horizons.}
	\label{nobh}
\end{figure*}

\begin{figure*}
	\begin{center}
	\begin{tabular}{c c}
		\includegraphics[scale=0.67]{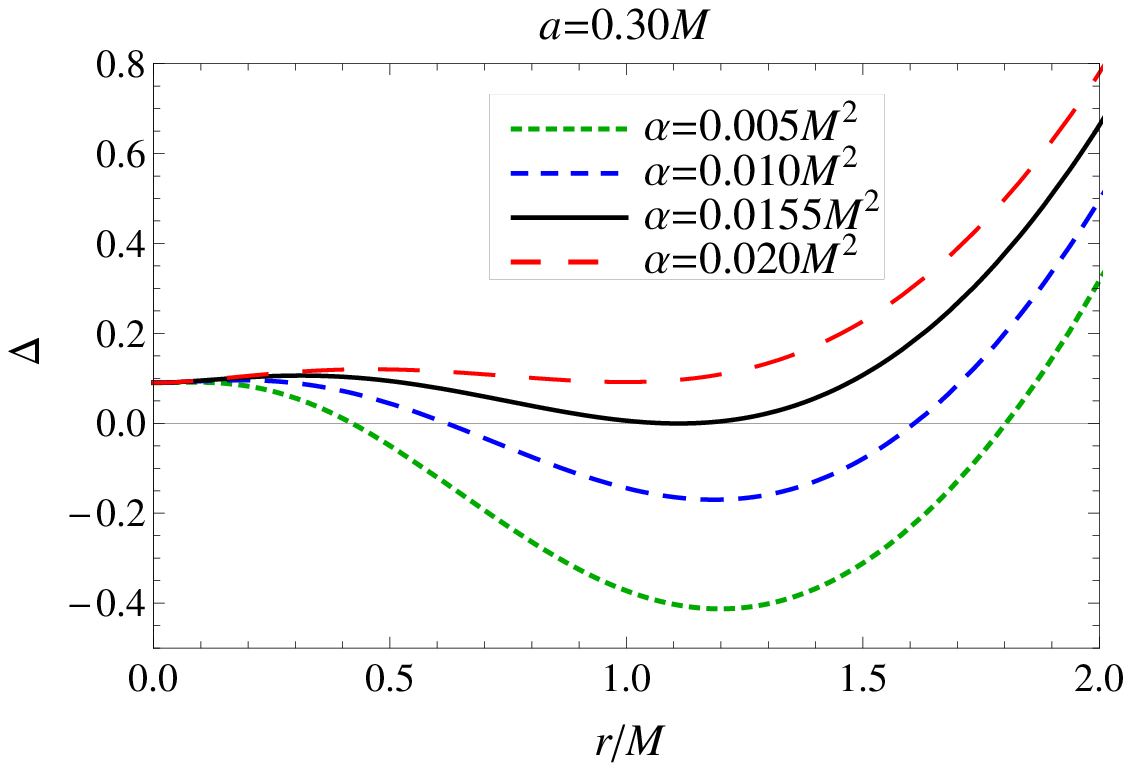}&
		\includegraphics[scale=0.67]{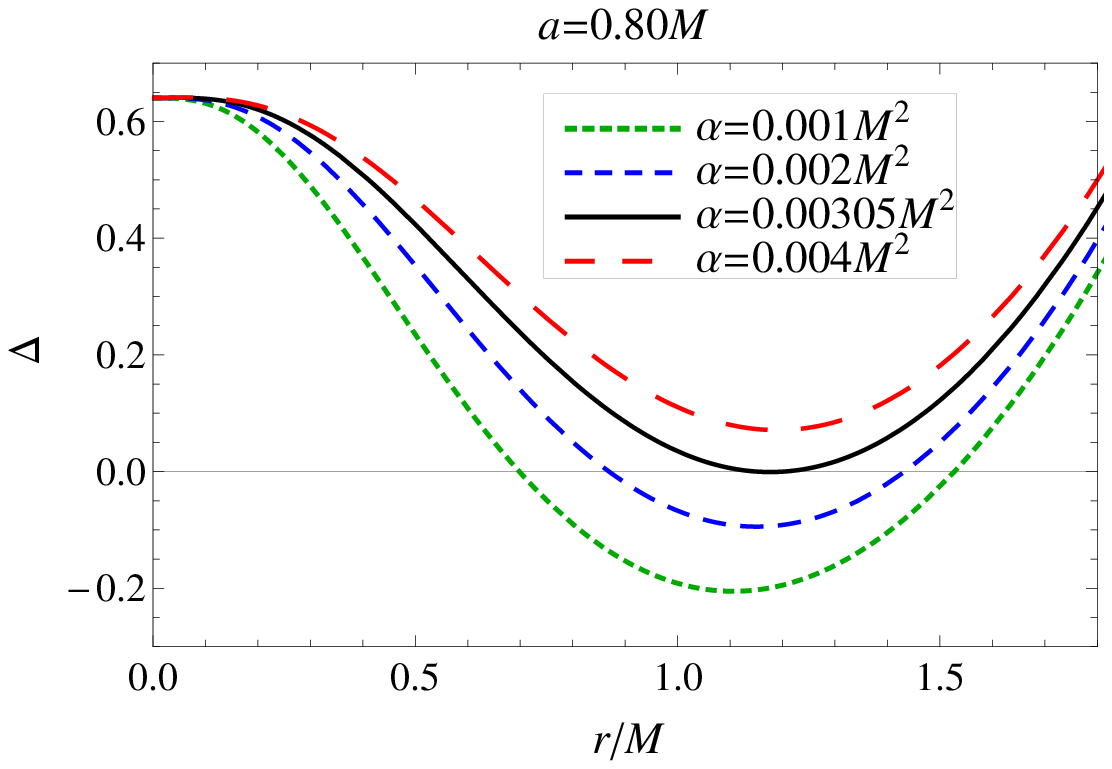}\\
		\includegraphics[scale=0.67]{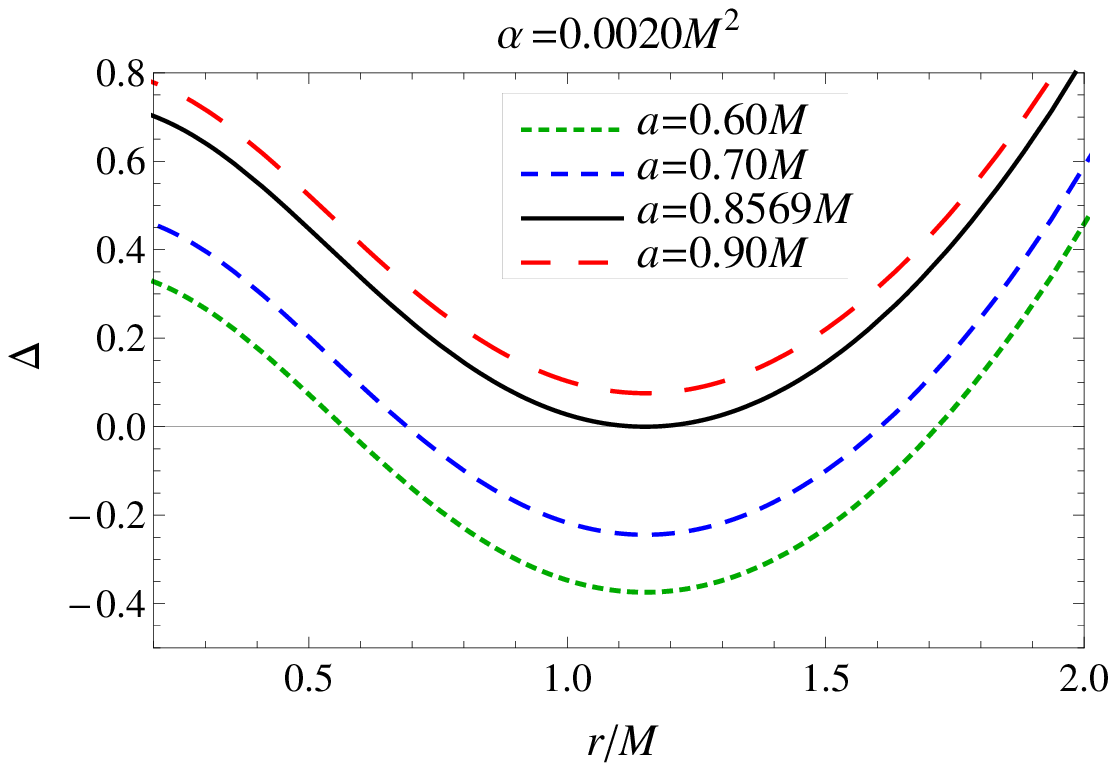}&
		\includegraphics[scale=0.67]{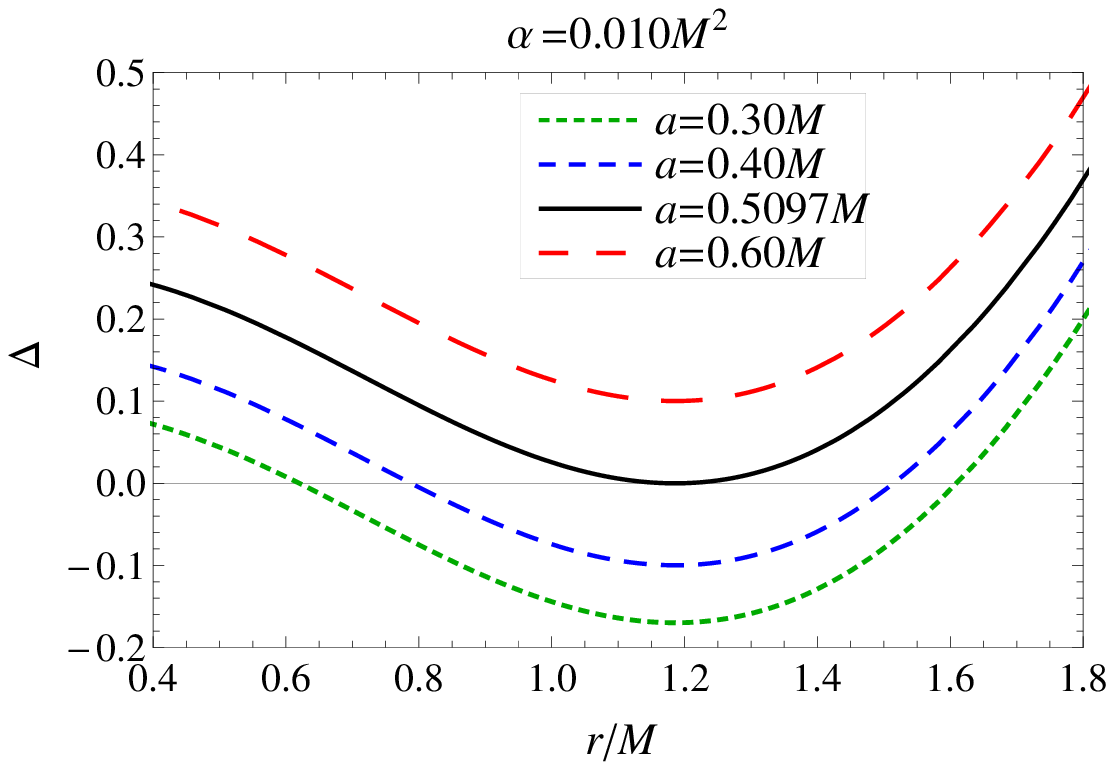}
	\end{tabular}
\end{center}
	\caption{The behavior of horizons with varying black hole parameters $a$ and $\alpha$. Black solid line corresponds to the extremal black hole with degenerate horizons.}
	\label{Horizonfig}
\end{figure*}

\begin{figure*}
	\includegraphics[scale=0.9]{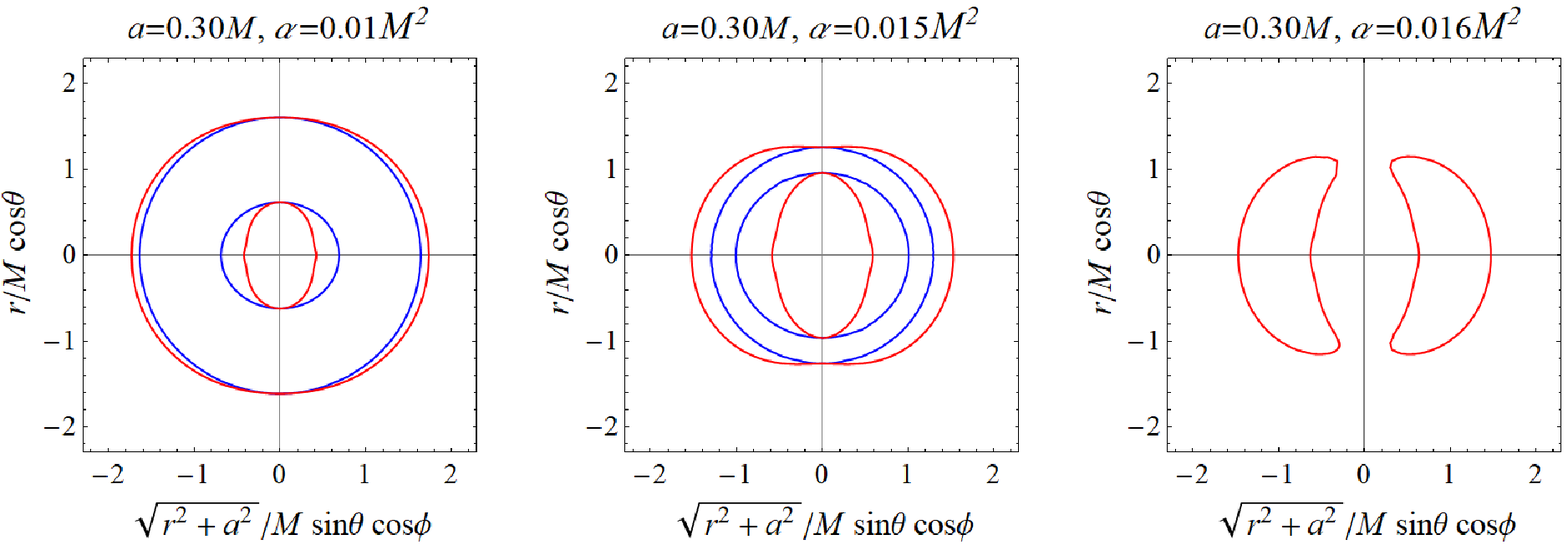}\\
	\includegraphics[scale=0.9]{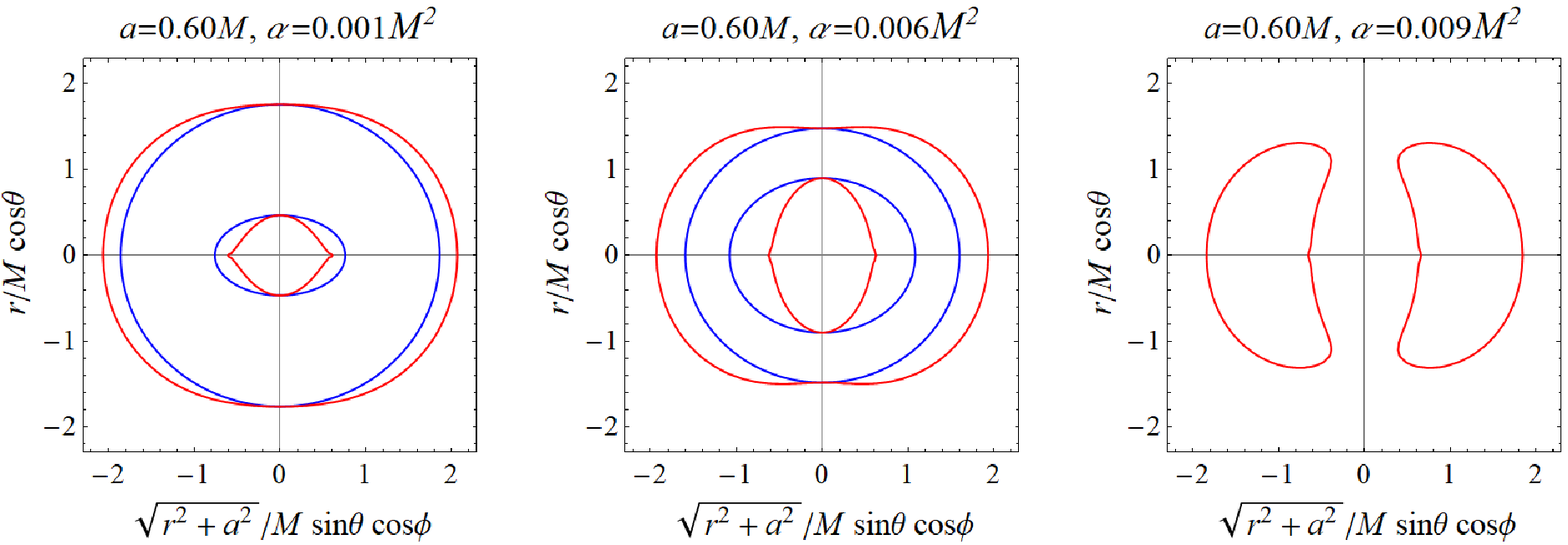}\\
	\includegraphics[scale=0.9]{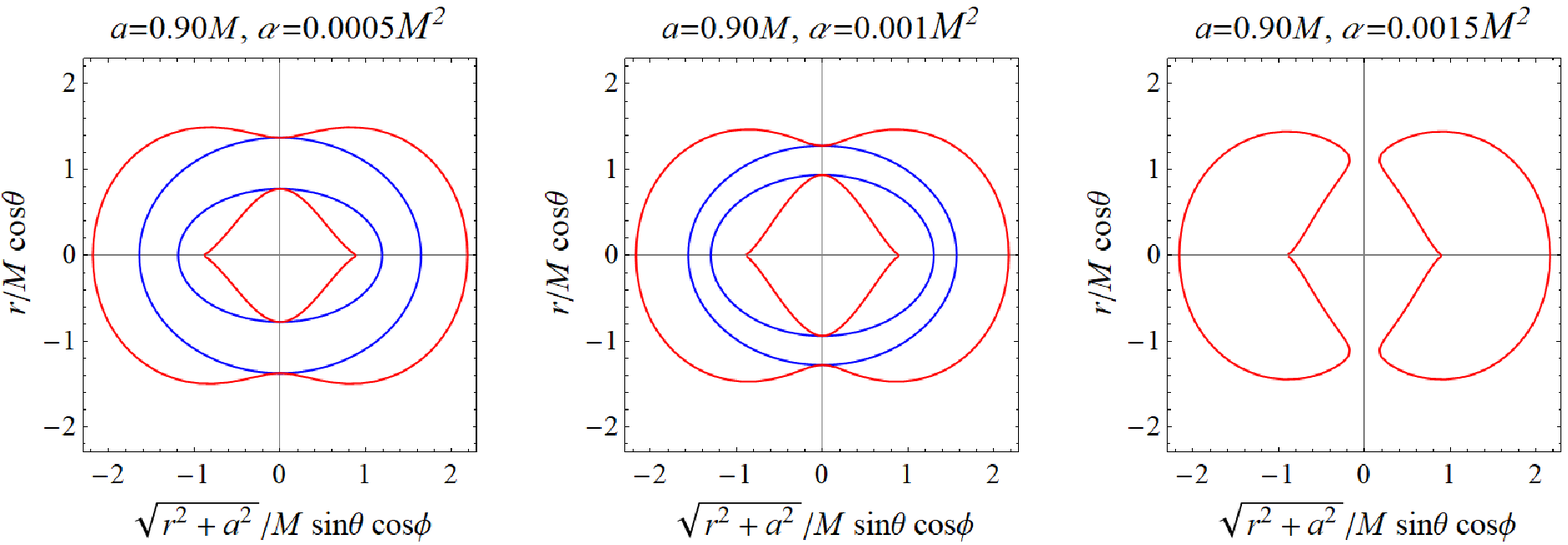}
	\caption{The cross-section of event horizon (outer blue line), SLS (outer red line) and ergoregion (region between event horizon and SLS) for different values of parameters $a$ and $\alpha$.} \label{Ergo}
\end{figure*}

Next, we investigate the nature of the metric (\ref{rotbhr}) to show that its properties are similar to the ones of the Kerr black hole. In particular, we compute the horizon-like surfaces with the aim to discuss the effect of the GB coupling parameter $\alpha$ on the structure of such surfaces. The horizons of rotating EGB black hole can be identified as the zeros of 
\begin{equation}
g^{\mu\nu}\partial_{\mu}r\partial_{\nu}r=g^{rr}=\Delta=0,\label{horizon}
\end{equation} 
which is also the coordinate singularity of the metric (\ref{rotbhr}). Numerical analysis reveals that depending on the values of $M, a$ and $\alpha$, eq.~(\ref{horizon}) can have maximum two distinct real positive roots or degenerate roots, or no-real positive roots, which respectively, correspond to the nonextremal black holes, extremal black holes, and no-black holes configurations for metric (\ref{rotbhr}). The existence condition of the horizons gives a bound on the black hole parameters $a$ and $\alpha$. In figure~(\ref{nobh}), the parameter space $(a,\alpha)$ is shown, for the parameters values within the gray region, metric (\ref{rotbhr}) admits two distinct roots, whereas for those outside no horizons exist and metric (\ref{rotbhr}) corresponds to the no-black hole spacetime. For the values of the parameters lying on the black solid line, metric (\ref{rotbhr}) admits degenerate roots and called the extremal black hole. The two real positive roots of eq.~(\ref{horizon}) are identified as the inner Cauchy horizon ($r_-$) and the outer event horizon ($r_+$) radii, such that $r_-\leq r_+$ (cf. figure~ \ref{Horizonfig}).
For the non-rotating black hole ($a=0$), eq.~(\ref{horizon}) admits solutions
\begin{equation}
r_{\pm}=M\pm \sqrt{M^2-16\pi\alpha }.
\end{equation}
The behavior of the horizon radii $r_{\pm}$ with varying spin parameter $a$ and GB coupling parameter $\alpha$ is shown in figure~\ref{Horizonfig}. For a fixed value of $a$, the event horizon radius $r_+$ decreases, while the Cauchy horizon radius $r_-$ increases with increasing $\alpha$. However, for a given value of $a$, there exist an extremal value of $\alpha$, i.e., $\alpha=\alpha_E$, for which both horizons coincide $r_-=r_+$, such that for $\alpha<\alpha_E$, horizon radii are $r_-\neq <r_+$ (cf. figure~(\ref{Horizonfig})). Further, the presence of the GB coupling reduces the horizon size, as for the fixed values of $M$ and $a$, the event horizon radii for the rotating EGB black holes are smaller as compared to that for the Kerr black hole (cf. figure~\ref{Horizonfig}). 
\begin{figure*}
	\begin{tabular}{c c}
		\includegraphics[scale=0.67]{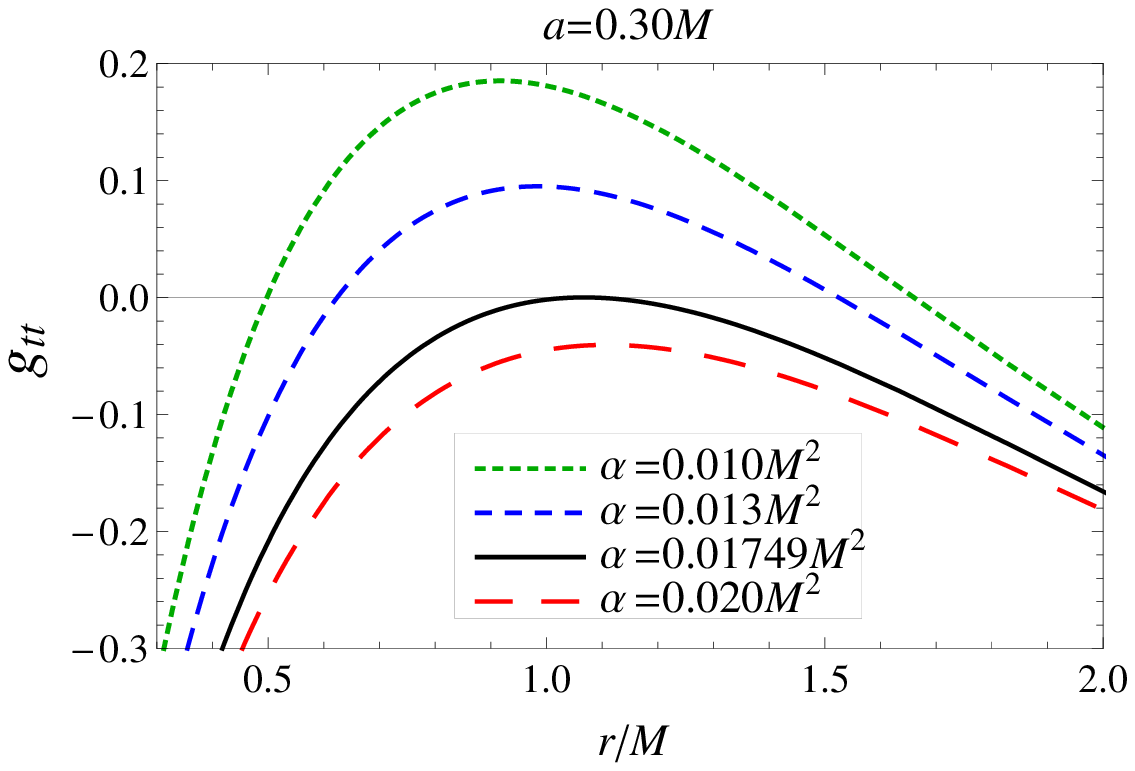}&
		\includegraphics[scale=0.67]{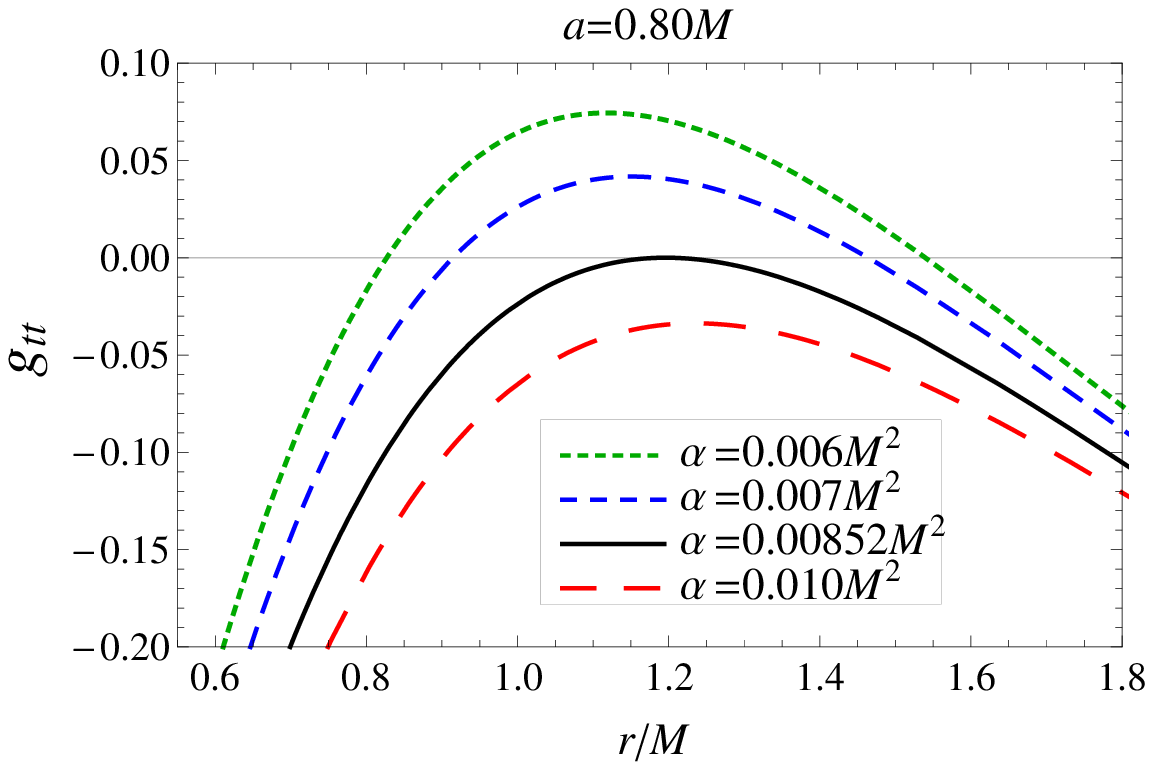}\\
		\includegraphics[scale=0.67]{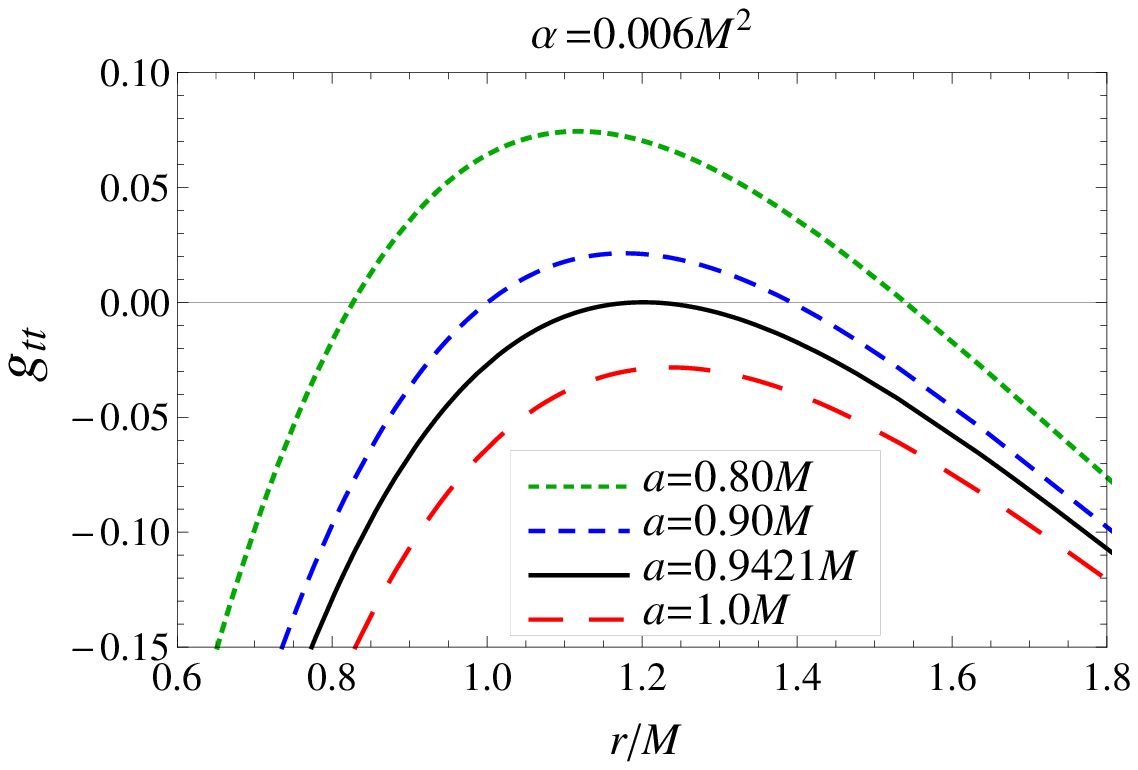}&
		\includegraphics[scale=0.67]{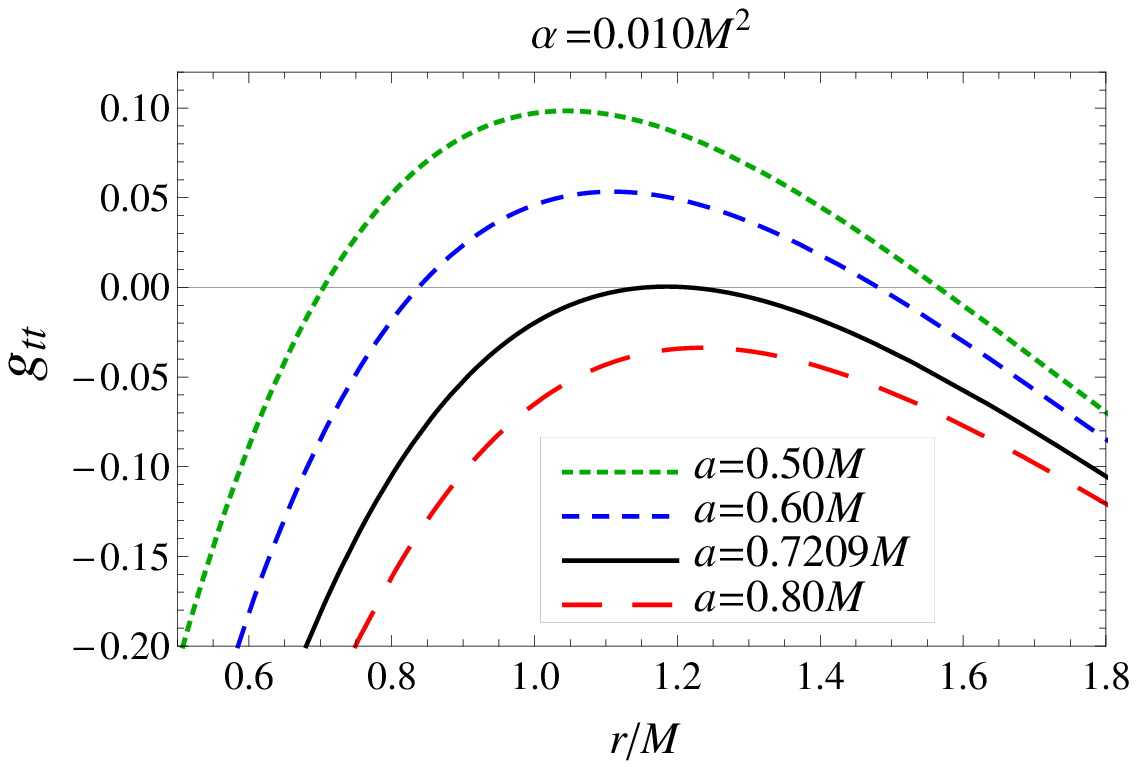}
	\end{tabular}
	\caption{The behavior of SLS with varying parameters $a, \alpha$, and $\theta=\pi/4$. The Black solid curve in each plot corresponds to the degenerate SLS.} \label{gtt}
\end{figure*}
The stationary observers, having zero angular momentum with respect to the distant observer at spatial infinity, outside the event horizon of the rotating black hole spacetime, can rotate along with the black hole rotation due to the frame-dragging effect \cite{Chandrasekhar:1992}. The angular velocity $\omega$ of the rotation reads
\begin{equation}
\omega=\frac{d\phi}{dt}=-\frac{g_{t\phi}}{g_{\phi\phi}}=\frac{a(r^2+a^2 - \Delta)} {\left[(r^2+a^2)^2-\Delta a^2\sin^2\theta\right]},
\end{equation}
$\omega$ monotonically increase with decreasing $r$ and eventually takes the maximum value at $r=r_+$, and reads as
\begin{equation}
\Omega=\left.\omega\right|_{r=r_+}=\frac{a} {(r_+^2+a^2)}=\frac{32\pi \alpha a}{r_+^4\left(-1+ \Big( 1+ \frac{128 \pi\alpha  M}{r_+^3} \Big)^{1/2}\right)}\label{angvelocity},
\end{equation}
which can  be identified as the black hole angular velocity  $\Omega$. Though stationary observers can exist outside the event horizons, static observers, following the worldlines of timelike Killing vector $\eta^{\mu}_{(t)}$, can exist only outside the static limit surface (SLS) defined by $\eta^{\mu}_{(t)}\eta_{\mu(t)}=g_{tt}=0$ \cite{Chandrasekhar:1992}. The radii of SLS are determined by the zeros of
\begin{equation}
r^2+a^2\cos^2\theta+\frac{r^4}{32\pi\alpha }\left(1- \Big( 1+ \frac{128 \pi\alpha  M}{r^3} \Big)^{1/2}\right)=0,\label{gtteq}
\end{equation}
which apart from black hole parameters also depends on $\theta$ and coincides with the event horizon only at the poles. Equation (\ref{gtteq}) is solved numerically and the two real positive roots, corresponding to the two SLS, are shown with varying $a$ and $\alpha$ in figure~\ref{gtt}. The radii of the outer SLS decreases with individually increasing $\alpha$ and $a$. For a fixed value of $a$, there exist a particular value of $\alpha$ for which the two SLS get coincide. However, these extremal values of $\alpha$ for $\theta\neq \pi/2$ are different from those for the degenerate horizons. For fixed values of $M$ and $a$, the SLS radii for the rotating EGB black holes are smaller as compared to the Kerr black hole values. The region between the SLS and the event horizon is termed as the ergoregion. It has been shown that it is possible, at least theoretically via Penrose process \cite{pen}, to extract energy from the ergosphere of the black hole as the region that lies outside of a black hole. In figure~\ref{Ergo}, we plotted the ergoregion of rotating EGB black holes, and it is evident that the size increase with $\alpha$. It will be interesting to see how the GB parameter $\alpha$ makes an impact on energy extraction. 
The surface gravity of the rotating EGB black hole takes the following form 
\begin{eqnarray}
\kappa&=&\left.\frac{\Delta'(r)}{2(r^2+a^2)}\right|_{r=r_+},\nonumber\\
&=&\frac{1}{2(r_+^2+a^2)}\left[2r_++\frac{r_+^3}{8\pi \alpha}-\frac{10M+\frac{r_+^3}{8\pi\alpha}}{\sqrt{\Big( 1+ \frac{128M \pi\alpha}{r_+^3} \Big)}}\right]
\end{eqnarray}
which in the limit of $\alpha \to 0$, reduces to
\begin{equation}
\kappa=\frac{r_+-M}{r_+^2+a^2},
\end{equation}
and that corresponds to the Kerr black hole surface gravity \cite{Frolov:1998wf}.

\section{Black hole shadow}\label{sec3}
The null geodesics describing the photon orbits around the black hole are especially interesting because of their observational importance in probing the gravitational impact of the black holes on the surrounding radiation. Photons originating from the light source behind the black hole arrive in the vicinity of the event horizon, and a part of it falls inside the horizon while another part scattered away to infinity. This along with the strong gravitational lensing results in the optical appearance of the black hole, namely, the black hole shadow encircled by the bright photon ring \cite{Bardeen,Synge:1966,Luminet:1979,CT}. The study of black hole shadow was led by the seminal work by Synge \cite{Synge:1966} and Luminet \cite{Luminet:1979}, who gave the formula to calculate the angular radius of the photon captured region around the Schwarzschild black hole by identifying the diverging light deflection angle. Later, Bardeen \cite{Bardeen} in his pioneering work studied the shadow of the Kerr black hole in a luminous background and shown that the spin would cause the shape of shadows distorted. The photon ring, encompassing the black hole shadow, explicitly depends on the spacetime geometry and thus its shape and size is a potential tool to determine the black hole parameters and to reveal the valuable information regarding the near-horizon field features of gravity. Over the past decades, a flurry of activities in the analytical investigations, observational studies and numerical simulation of shadows for large varieties of black holes have been reported \cite{De1,Grenzebach1}. Black hole shadows have also been investigated in the context of black hole parameter estimations and in testing theories of gravity \cite{Kumar:2018ple,Kramer1:2004hd}. \\
We use the Hamilton Jacobi equation and Carter's separable method \cite{Carter:1968rr} to determine the geodesics motion in the rotating black hole spacetime (\ref{rotbhr}). The four integrals of motions, namely, the particle rest mass $m_0$, energy $\cal E$, axial angular momentum $\cal L$ and the Carter constant $\mathcal{K}$ related to the latitudinal motion of the test particle, completely describe the geodesics equations of motion in the first-order differential form  \cite{Carter:1968rr,Chandrasekhar:1992}
\begin{eqnarray}
\Sigma \frac{dt}{d\tau}&=&\frac{r^2+a^2}{\Delta}\left({\cal E}(r^2+a^2)-a{\cal L}\right)  -a(a{\cal E}\sin^2\theta-{\mathcal {L}}) ,\label{tuch}\\
\Sigma \frac{dr}{d\tau}&=&\pm\sqrt{\mathcal{R}(r)} ,\label{r}\\
\Sigma \frac{d\theta}{d\tau}&=&\pm\sqrt{\Theta(\theta)} ,\label{th}\\
\Sigma \frac{d\phi}{d\tau}&=&\frac{a}{\Delta}\left({\cal E}(r^2+a^2)-a{\cal L}\right)-\left(a{\cal E}-\frac{{\cal L}}{\sin^2\theta}\right), \label{phiuch}
\end{eqnarray}
where $\tau$ is the affine parameter along the geodesics and 
\begin{eqnarray}\label{06}
\mathcal{R}(r)&=&\left((r^2+a^2){\cal E}-a{\cal L}\right)^2-\Delta ((a{\cal E}-{\cal L})^2+{\cal K}),\quad \\ 
\Theta(\theta)&=&{\cal K}-\left(\frac{{\cal L}^2}{\sin^2\theta}-a^2 {\cal E}^2\right)\cos^2\theta.\label{theta0}
\end{eqnarray}
The separable constant ${\cal K}$ is related to the Carter's constant of motion $\mathcal{Q}=\mathcal{K}+(a\mathcal{E}-\mathcal{L})^2$ \cite{Chandrasekhar:1992,Carter:1968rr}. Let define the two dimensionless impact parameters,
\begin{equation}
\eta\equiv{\cal K}/\mathcal{E}^2,\qquad \xi\equiv\mathcal{L}/\mathcal{E},
\end{equation} 
which parameterize the null geodesics. Unstable photon orbits, at constant Boyer-Lindquist coordinate $r_p$, are determined by the vanishing radial potential and its radial derivative 
\begin{equation}
\left.\mathcal{R}\right|_{(r=r_p)}=\left.\frac{\partial \mathcal{R}}{\partial r}\right|_{(r=r_p)}=0\;\; \text{and} \quad \left.\frac{\partial^2 \mathcal{R}}{\partial r^2}\right|_{(r=r_p)}> 0.\label{pot}
\end{equation}
\begin{figure*}
	\begin{tabular}{c c c}
		\includegraphics[scale=0.8]{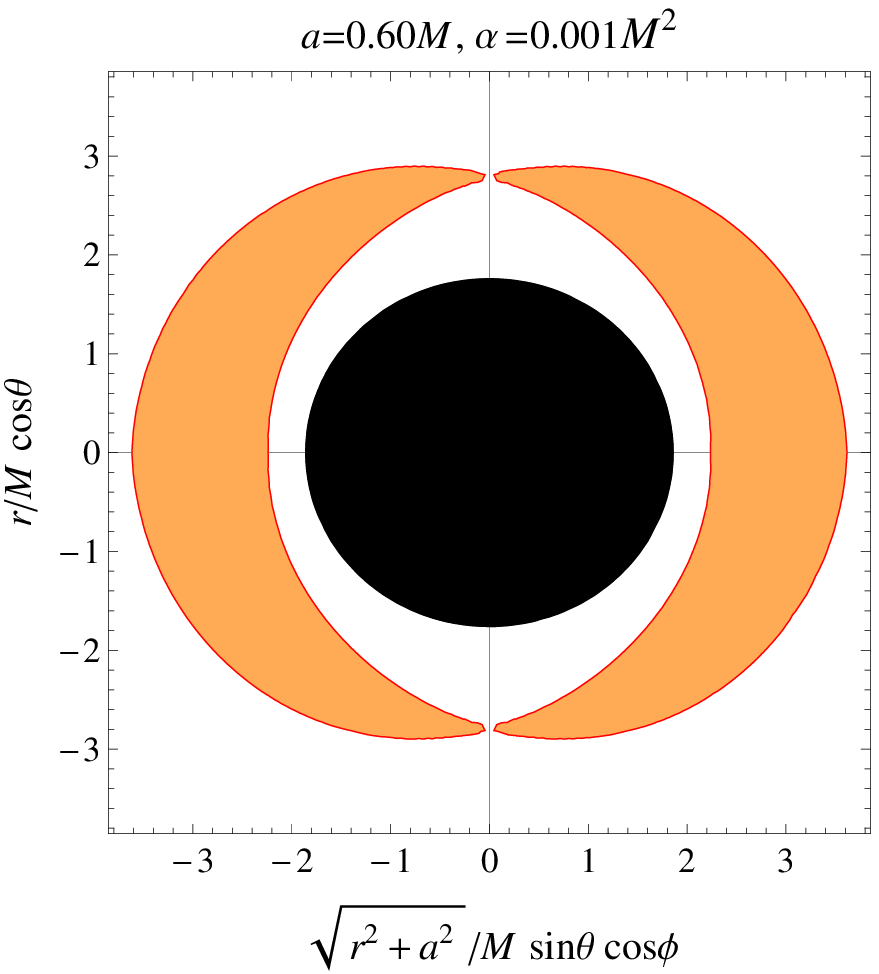}&
		\includegraphics[scale=0.8]{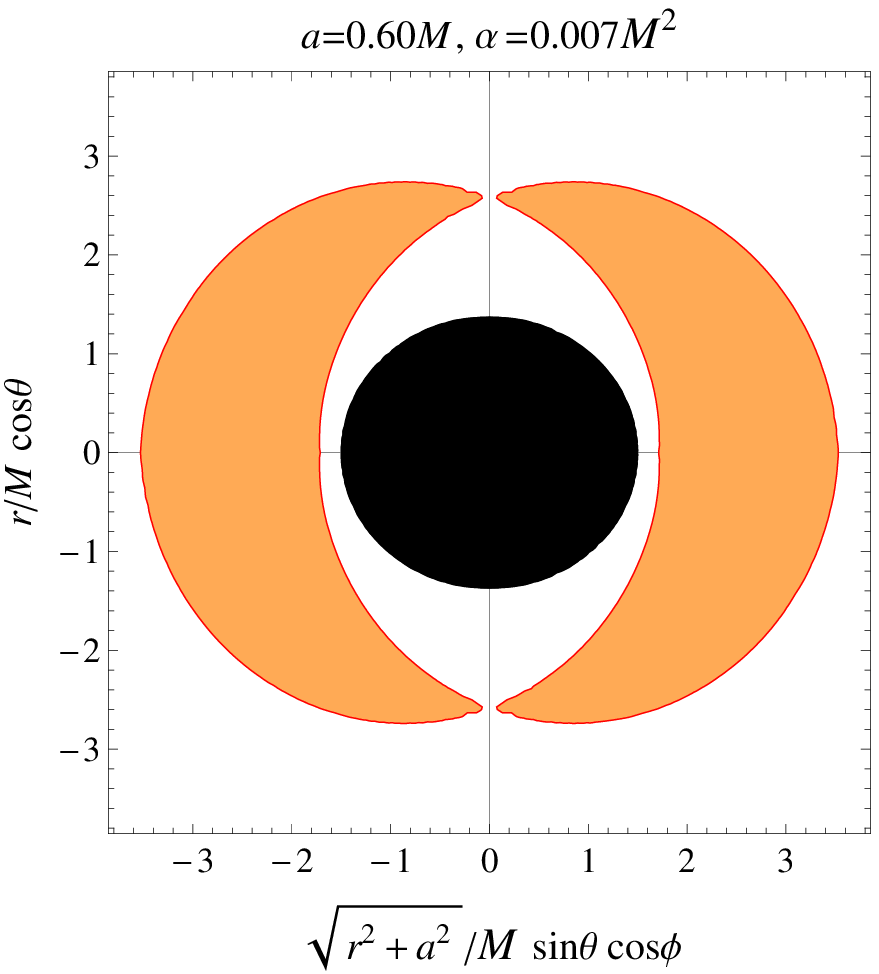}
	\end{tabular}
	\caption{The photon region (shaded orange region) structure around the rotating EGB black hole (shaded black disk).}
	\label{ph}
\end{figure*}
Solving eq.~(\ref{pot}) yields the pair of critical values of impact parameters ($\eta_c, \xi_c$) for the unstable photon orbits
\begin{eqnarray}
\eta_c&=& \frac{r_p^2 \Bigr(-16 \Delta(r_p)^2 - r_p^2 \Delta'(r_p)^2 + 8 \Delta(r_p)\Big (2 a^2 + r_p \Delta '(r_p)\Big)\Bigl)}{a^2 \Delta '(r_p)^2},
 \nonumber\\
\xi_c&=&\frac{\Big(r_p^2+a^2\Big)\Delta '(r_p)-4r_p \Delta(r_p)}{a\Delta '(r_p)},\label{impactparameter}
\end{eqnarray}
which effectively separate the captured orbits
from the scattered one. Planer circular orbits are possible only at the equatorial plane ($\theta=\pi/2$) for $\eta_c=0$, whereas, generic spherical photon orbits exist for $\eta_c>0$. The radii $r_p^{\pm}$ of co-rotating and counter-rotating circular orbits can be identified as the real positive zeros of $\eta_c=0$. These spherical photons orbits construct the photon regions, which are determined by Eqs.~(\ref{theta0}) and (\ref{impactparameter}), and given by ($\Theta(\theta)\geq 0$)
\begin{equation}
(4r_p\Delta(r_p)-\Delta'(r_p)\Sigma)^2\leq 16a^2r_p^2\Delta(r_p)\,\sin^2\theta,
\end{equation}
the gravitationally lensed image of this photon region corresponds to the black hole shadow (cf. figure~\ref{ph}). Figure \ref{ph} infers that the photon region size increases with the GB coupling parameter $\alpha$. Indeed rotating black holes have two distinct photon regions, one inside the Cauchy horizon and another outside the event horizon \cite{Grenzebach}. Black hole image observations morphology relies on photons that can reach the observer, therefore, we look only for the unstable photons orbits lying outside the event horizon, i.e., $r_p>r_+$. We consider an observer residing far away from the black hole ($r_o, \theta_o$) so that the observer's neighborhood can be taken as asymptotically flat. The observer can pick a Cartesian coordinate system centered at the black hole, such that the projection of the spherical photon orbits on the celestial plane delineates a closed curve parameterized by the celestial coordinates ($X, Y$). Back tracing a photon trajectory from the observer's position to the celestial plane marks a point on the image plane, described as follow  \cite{Bardeen,CT}
\begin{eqnarray}
X&=&-r_o\frac{p^{(\phi)}}{p^{(t)}}=\lim_{r_o\rightarrow\infty}\left(-r_o^2 \sin{\theta_o}\frac{d\phi}{d{r}}\right)\nonumber\\ Y&=&r_o\frac{p^{(\theta)}}{p^{(t)}}=\lim_{r_o\rightarrow\infty}\left(r_o^2\frac{d\theta}{dr}\right),\label{Celestial}
\end{eqnarray} 
where $p^{(\mu)}$ are the tetrad components of the photon four-momentum with respect to a locally nonrotating reference frame. Equation~(\ref{Celestial}) can be further simplified in terms of impact parameters as follow \cite{Bardeen,CT}
\begin{eqnarray}
X&=&-\xi_c\csc\theta_o\xrightarrow{\theta_o=\pi/2} -\xi_c,\nonumber\\
Y&=&\pm\sqrt{\eta_c+a^2\cos^2\theta_o-\xi_c^2\cot^2\theta_o}\xrightarrow{\theta_o=\pi/2}\pm\sqrt{\eta_c}.\label{pt}
\end{eqnarray} 
The parametric curve $Y$ vs $X$ delineates the rotating EGB black hole shadow.

\begin{figure*}
\begin{center}
	\begin{tabular}{c c}
		\includegraphics[scale=0.85]{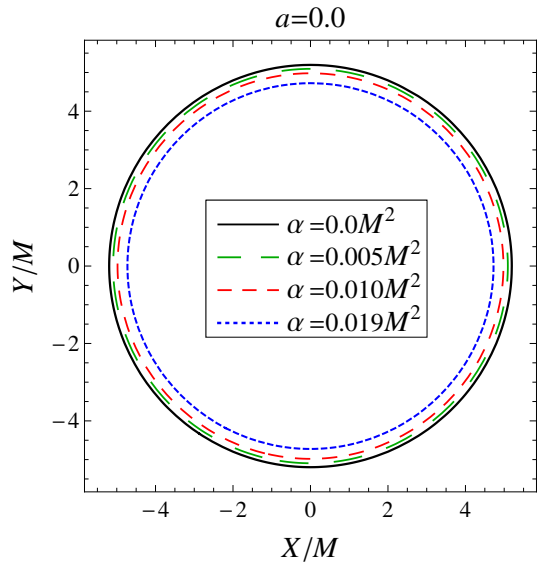}
	\end{tabular}
\end{center}
	\caption{Non-rotating EGB black hole shadows with varying coupling parameter $\alpha$. Black solid line corresponds to the Schwarzschild black hole shadow. }
	\label{shadowNR}
\end{figure*}
\begin{figure*}
	\begin{tabular}{c c}
		\includegraphics[scale=0.9]{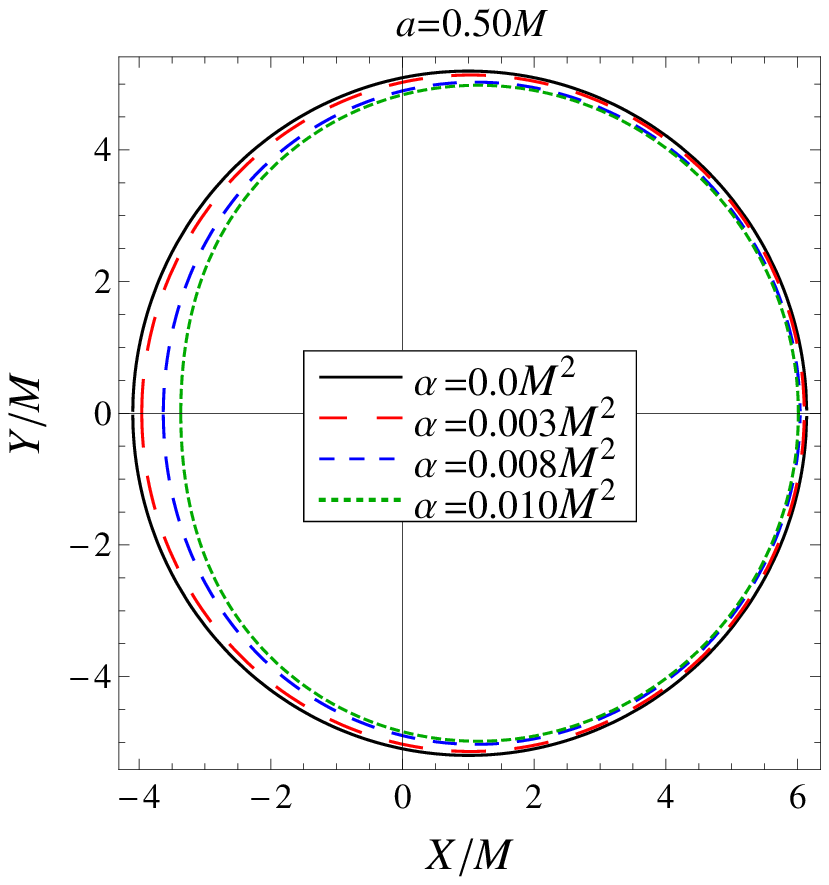}&
		\includegraphics[scale=0.9]{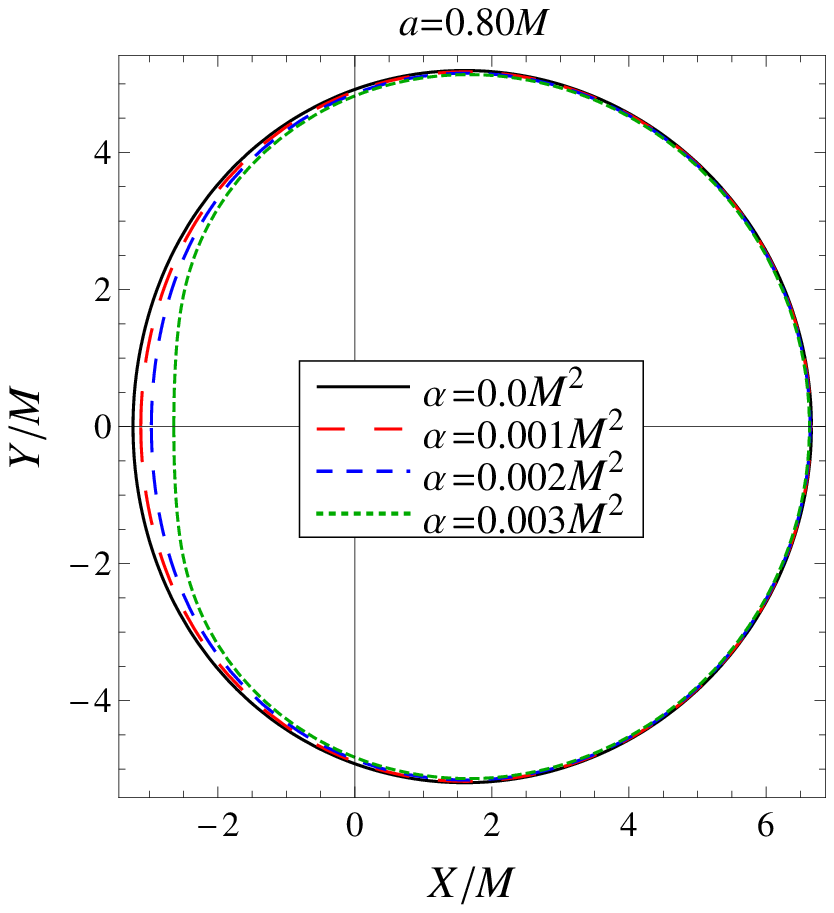}\\
		\includegraphics[scale=0.9]{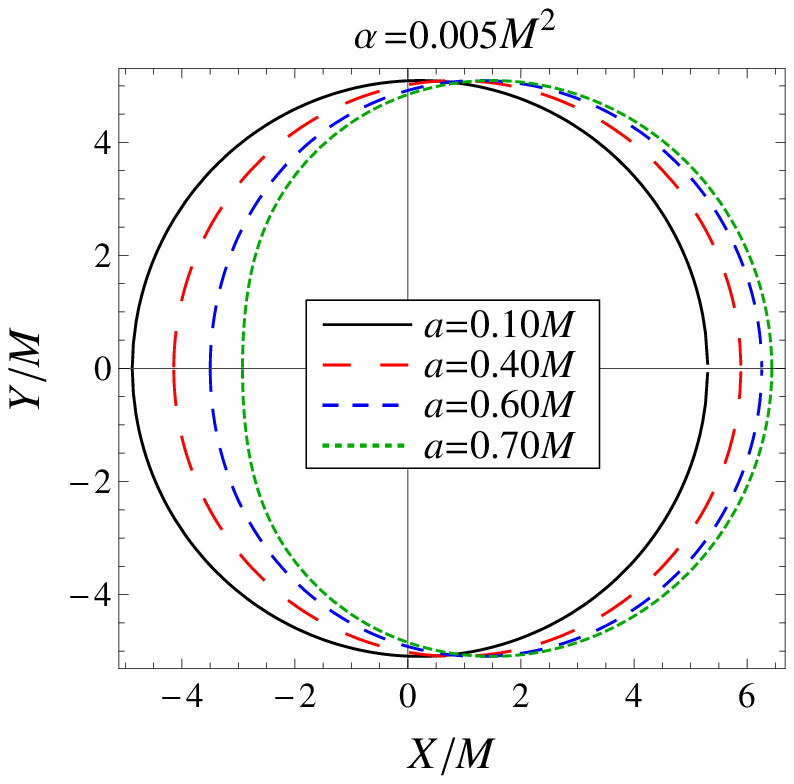}&
		\includegraphics[scale=0.9]{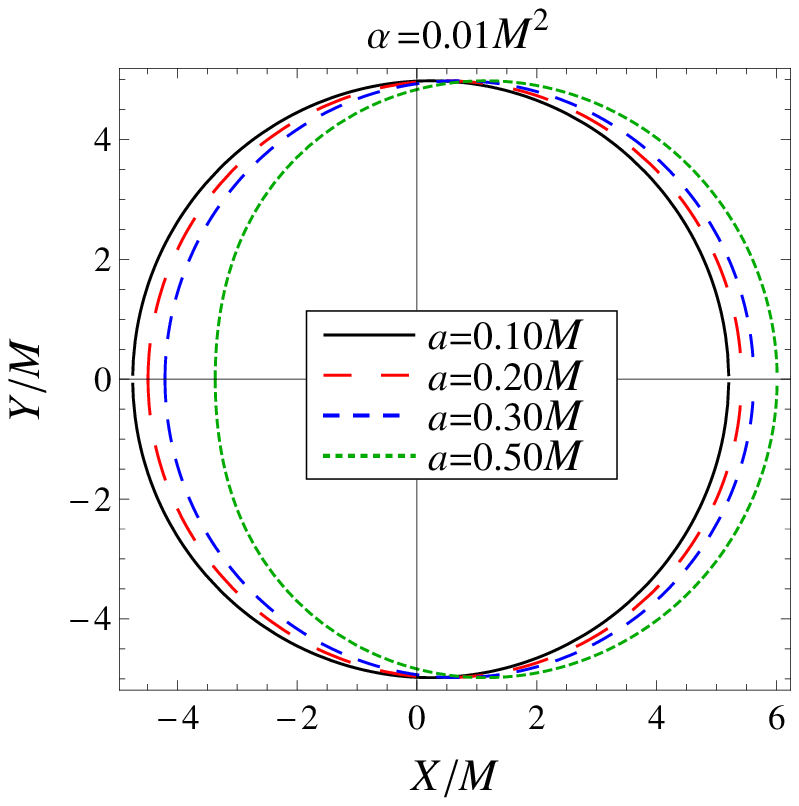}
	\end{tabular}
		\caption{Plot showing the rotating EGB black holes shadows with varying parameters $a$ and $\alpha$. Solid black curves in the upper panel are for the Kerr black holes. }
\label{shadow}
\end{figure*}
\begin{figure*}
	\begin{tabular}{c c c}
		\includegraphics[scale=0.8]{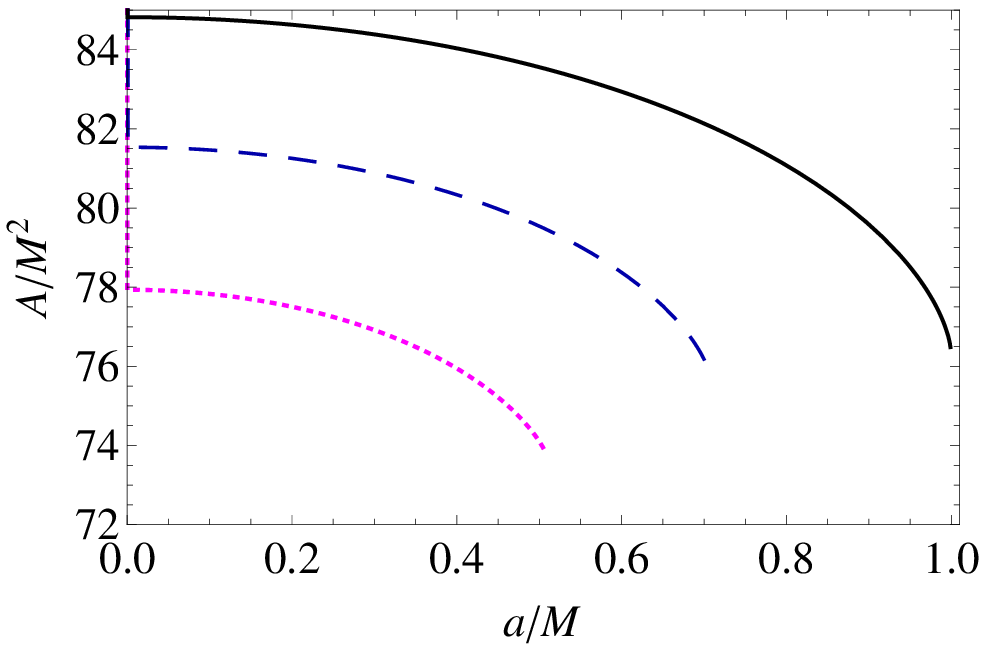}&
		\includegraphics[scale=0.8]{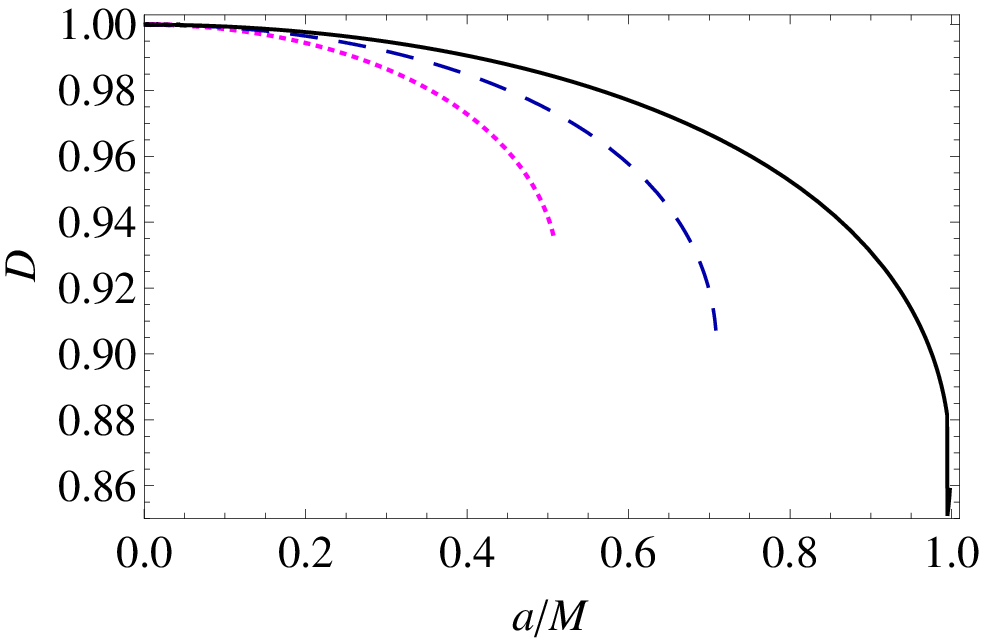}
	\end{tabular}
	\caption{The shadow area $A$ and oblateness observables $D$ vs $a$ for the rotating EGB black holes, (solid Black curve) for the Kerr black hole $\alpha=0.0M^2$, (dashed Blue curve) for $\alpha=0.005M^2$, and (dotted Magenta curve) for $\alpha=0.01M^2$.}
	\label{obs1}
\end{figure*}
\begin{figure*}
	\begin{tabular}{c c c}
		\includegraphics[scale=0.8]{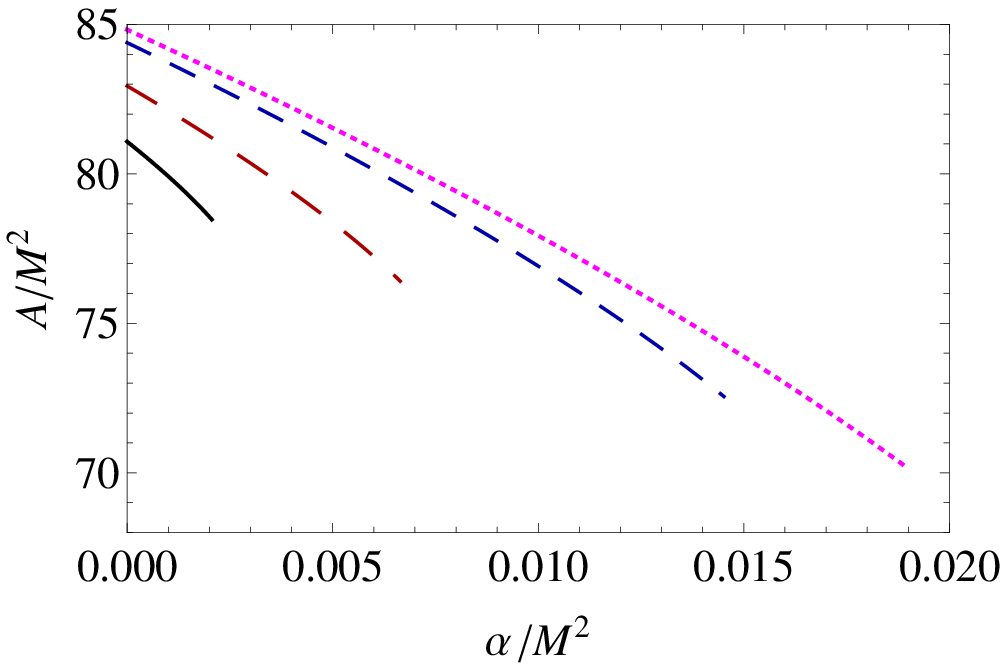}&
		\includegraphics[scale=0.8]{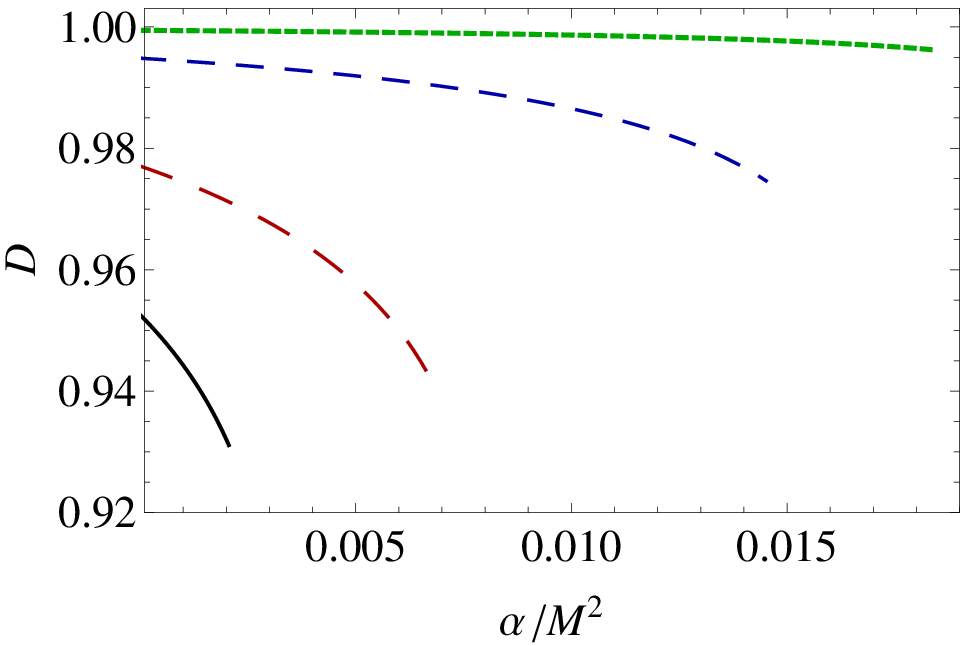}
	\end{tabular}
	\caption{The shadow area $A$ and oblateness observables $D$ vs $\alpha$ for the rotating EGB black holes, (dotted Magenta curve) for $a=0.0M$, (dotted Green curve) for $a=0.1M$, (dashed Blue curve) for $a=0.3M$, (long-dashed Brown curve) for $a=0.6M$, and (solid Black curve) for $a=0.8M$.}
	\label{obs2}
\end{figure*}
\begin{figure*}
\begin{center}
	\includegraphics[scale=0.9]{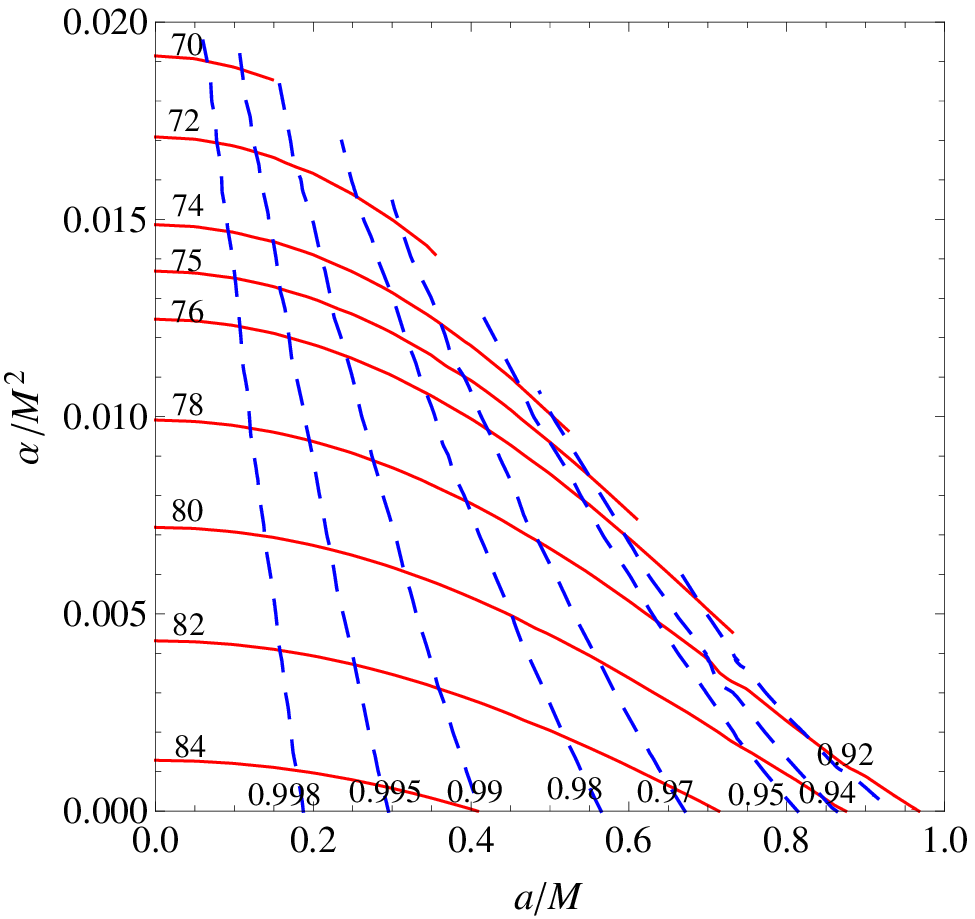}
\end{center}
	\caption{Contour plots of the observables $A$ and $D$ in the plane $(a, \alpha)$ for the rotating EGB black holes. Each curve is labeled with the corresponding values of $A$ and  $D$. Solid red curves correspond to the $A$, and dashed blue curves are for the oblateness parameter $D$. }
	\label{parameterestimation}
\end{figure*}
For the non-rotating case, eq.~(\ref{pt}) yield 
\begin{equation}
X^2+Y^2= \left.2 r_P \Bigr(r_p + \frac{4\Delta(r_p)(2r_p-\Delta'(r_p))}{\Delta'(r_p)^2}\Bigl)\right|_{a=0},\label{nonrot}
\end{equation}
which shows that the static spherically symmetric EGB black hole cast a perfectly circular shadow (cf. figure~\ref{shadowNR}). The size of the shadow decreases with increasing $\alpha$, such that the non-rotating EGB black holes shadows are smaller than the Schwarzschild black hole shadow as shown in figure~\ref{shadowNR}. Rotating EGB black hole shadows with varying $a$ and $\alpha$ are shown in figure~\ref{shadow}, which clearly infer that the presence of the GB coupling parameter has a profound influence on the apparent shape and size of the shadow. For non-zero values of $a$, the rotating black holes shadows are not perfect circles. The size and the amount of deviation from the circularity are measurable quantities and therefore, we introduce the shadow observables, namely, shadow area $A$ and oblateness $D$, for the characterization of shadows \cite{Kumar:2018ple,Tsupko:2017rdo}
\begin{equation}
A=2\int{Y(r_p) dX(r_p)}=2\int_{r_p^{-}}^{r_p^+}\left( Y(r_p) \frac{dX(r_p)}{dr_p}\right)dr_p,\label{Area}
\end{equation} 
\begin{equation}
D=\frac{X_r-X_l}{Y_t-Y_b},\label{Oblateness}
\end{equation}
where subscripts $r, l, t$ and $b$, respectively, stand for the right, left, top and bottom of the shadow boundary. Observables $A$ and $D$, respectively, characterize the shadow size and shape; for a perfectly circular shadow $D=1$.  Figure~\ref{obs1} shows the behavior of the shadow observables $A$ and $D$ with varying $a$ for different values of
$\alpha$. It is evident that the shadows of the rotating EGB black holes are smaller and more distorted as compared to the Kerr black hole shadow as depicted by the black solid curves in figure~\ref{obs1}. The shadow size decrease and the oblateness increases with increasing $a$. In figure~\ref{obs2}, the variation of the $A$ and $D$ with $\alpha$ is shown for various values of $a$. The shadow size monotonically decreases with $\alpha$. For the estimation of the rotating EGB black hole parameters, we plotted $A$ and $D$ as functions of $a$ and $\alpha$ in figure~\ref{parameterestimation}. Each curve corresponds to constant values of either $A$ or $D$. The point of intersection of the $A$ and $D$ curves gives the unique values of the black hole parameters $a$ and $\alpha$. In Table~\ref{tab1}, we summarize the extracted values of EGB black hole parameters from the known shadow observables. Hence, from figure~\ref{parameterestimation} and Table~\ref{tab1}, it is clear that for a given set of $4D$ EGB black hole shadow observables, $A$ and $D$, we can determine information about black hole spin and GB coupling parameter. 
\begin{table}[h!]
	\centering
	\begin{tabular}{ p{1.9cm} p{3cm} p{1.9cm} p{1.9cm} }
		\hline 	\hline
		\multicolumn{2}{l}{Shadow Observables }  & 	\multicolumn{2}{c}{Black Hole Parameters}\\
		\hline
		$A/M^2 $ &  $D$ &  $a/M$ & $\alpha/M^2$  \\
		\hline\hline
		82.0&   0.997153 &  0.19& 0.003974 \\
		\hline
		82.0&0.983341 &  0.471& 0.002288 \\
		\hline
		81.8517&	0.99058 &  0.346& 0.0034 \\
		\hline
		78.0 & 0.92429 & 0.81005 & 0.0022\\
		\hline
		76.92& 0.89092&0.910 & 0.0011\\
		\hline
		75.0&0.99342 & 0.18396& 0.0131 \\
		\hline
		72.0& 0.972212& 0.29955& 0.0150 \\
		\hline
		70.0 & 0.99802 & 0.070 & 0.019\\
		\hline
		69.8686&0.9959 &  0.099& 0.01899 \\
		\hline\hline
	\end{tabular}
	\caption{Estimated values of rotating EGB black hole parameters $a$ and $\alpha$ from known shadow observables $A$ and $D$.}\label{tab1}
\end{table}
\begin{figure*}
	\begin{tabular}{c c}
		\includegraphics[scale=0.7]{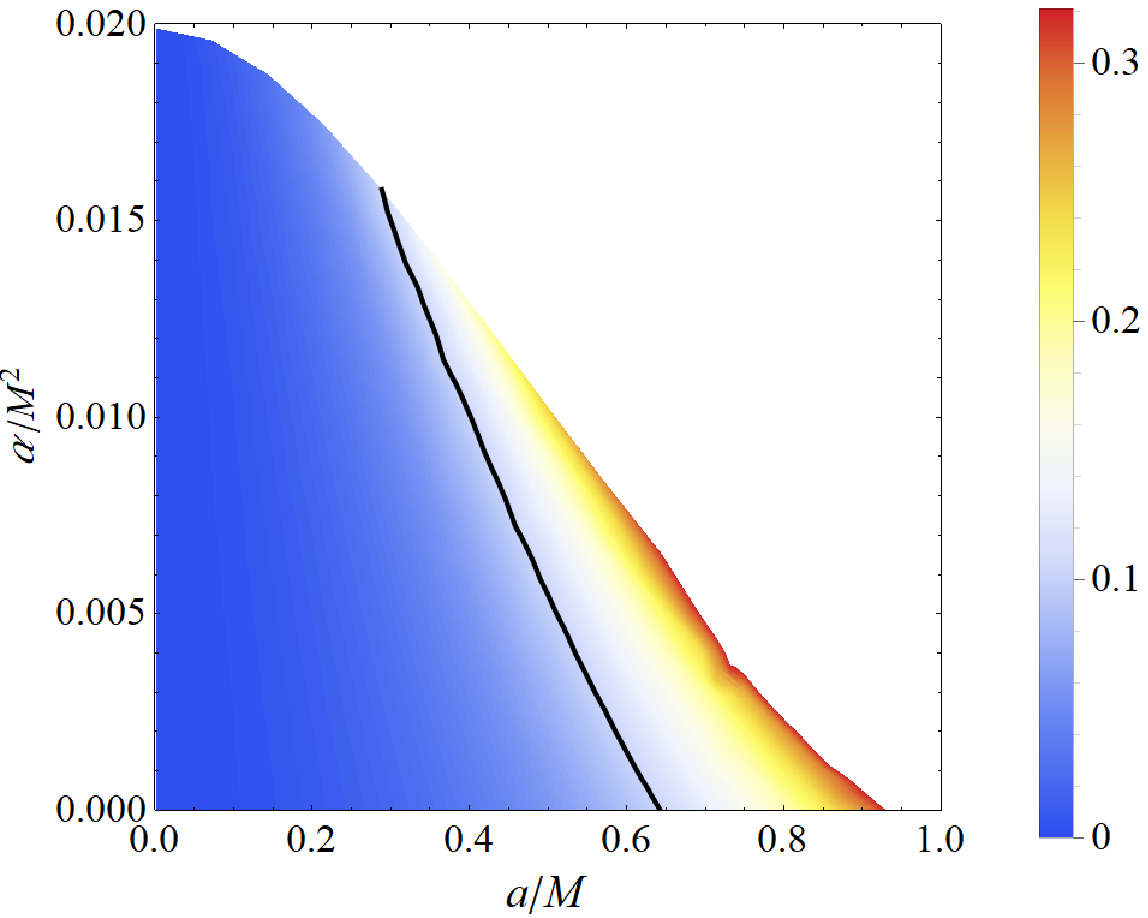}&
		\includegraphics[scale=0.7]{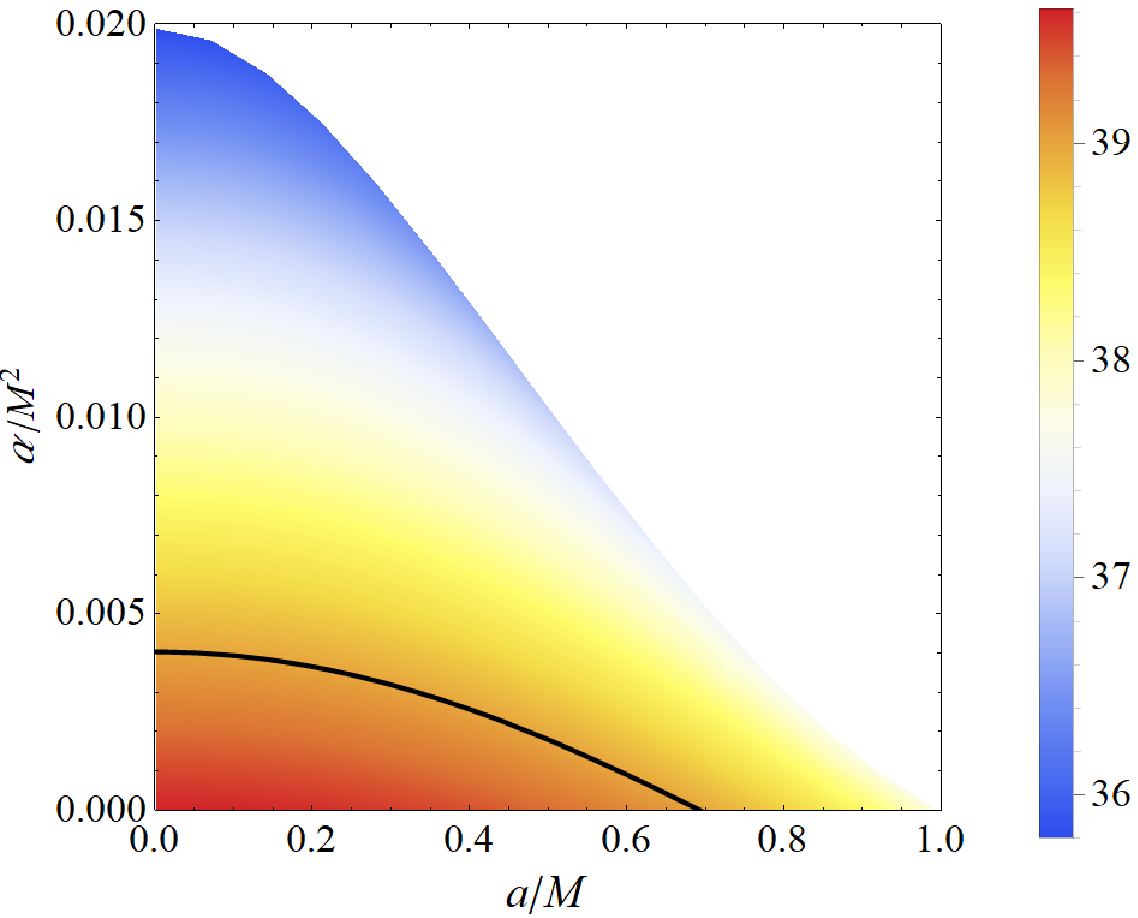}
	\end{tabular}
	\caption{Circularity deviation observable $\Delta C$ (left panel) and the angular diameter $\theta_d$ (right panel) as a function of ($a,\alpha$) for the rotating EGB black holes. Black solid lines correspond to the M87* black hole shadow bounds, namely $\Delta C=0.10$  and $\theta_d=39\mu$as within the $1\sigma$ region, such that the region above the black line is excluded by the EHT bounds.}
	\label{M87}
\end{figure*}
We can parameterize the shadow boundary also by the radial and angular coordinates ($R(\varphi),\varphi$) in a polar coordinate system with the origin at the shadow center ($X_C,Y_C$). Figure (\ref{shadow}) infers that the black hole shadow is symmetric under reflection around $Y=0$. However, it is not symmetric under reflections around the $X$ axis and is shifted from $X=0$. This ensure that the $X_C=(X_r - X_l)/2$, and $Y_C=0$. A point on the shadow boundary has a radial distance $R(\varphi)$ from the shadow center and subtends an angle $\varphi$ on the $X$ axis at the geometric center ($X_C,0$), which reads as
\begin{equation}
R(\varphi)= \sqrt{(X-X_C)^2+(Y-Y_C)^2},\;\ \varphi\equiv \tan^{-1}\left(\frac{Y}{X-X_C}\right)\nonumber.
\end{equation}
The shadow average radius $\bar{R}$ is defined as \cite{Johannsen:2010ru}
\begin{equation}
\bar{R}=\frac{1}{2\pi}\int_{0}^{2\pi} R(\varphi) d\varphi,
\end{equation}
We describe the circularity deviation $\Delta C$ as a measures of the root-mean-square deviation of $R(\varphi)$ from the shadow average radius  \cite{Johannsen:2010ru,Johannsen:2015qca}
\begin{equation}
\Delta C=2\sqrt{\frac{1}{2\pi}\int_0^{2\pi}\left(R(\varphi)-\bar{R}\right)^2d\varphi},
\end{equation}
$\Delta C$ quantifies the shadow deviation from a perfect circle, such that for a spherically symmetric black hole circular shadow, $\Delta C=0$. We can use our numerical results to compare the latest observation of the black hole shadow of M87*. The EHT Collaboration \cite{EHT} using the very large baseline interferometry technique has observed the shadow of M87* black hole residing at the center of nearby galaxy M87 \cite{Akiyama:2019cqa,Akiyama:2019eap}. Their measurement of the M87* black hole mass ($M=(6.5\pm 0.7) \times 10^9 M_{\odot}$) is consistent with the prior mass measurement using stellar dynamics ($M=(6.2\pm 0.8) \times 10^9 M_{\odot}$) but inconsistent with the gas dynamics measurement ($M=(3.2\pm 0.6) \times 10^9 M_{\odot}$). Though the observed shadow is found to be consistent with the general relativistic magnetohydrodynamics simulated images of the Kerr black hole as predicted by the general relativity, various Kerr modified black hole models in general relativity as well in modified gravities could not be completely ruled out currently \cite{Bambi1}. The measured circularity deviation, $\Delta C\leq 0.10$, for the M87* black hole shadow can be used to constrain the GB coupling parameter. $\Delta C$ is calculated for the rotating EGB black hole metric eq.~(\ref{rotbhr}) and plotted as a function of ($a,\alpha$) in Fig~\ref{M87}. The EHT bound for the M87* black hole shadow, shown as the black solid line, is used to constrain the $a$ and $\alpha$. It is clear that the $\Delta C$ merely puts a constrain on the EGB parameter $\alpha$.

Further, a far distant observer, at a distance $d$ from the black hole, measures the angular diameter $\theta_d$ for the black hole shadow 
\begin{equation}
\theta_d=2\frac{R_s}{d},\;\;\;\; R_s=\sqrt{A/\pi},
\end{equation}    
The inferred angular diameter for the M87* black hole shadow is $\theta_d=42\pm 3\, \mu$as \cite{Akiyama:2019cqa}. We calculated the angular diameter for the rotating EGB black hole shadow for $M=(6.5\pm 0.7) \times 10^9 M_{\odot}$, and $d=16.8\, M$pc, and plotted as a function of $a$ and $\alpha$, the region enclosed by the black solid line, $\theta_d=39\,\mu$as, falls within the $1\sigma$ region of the M87* shadow angular diameter. For the given mass and distance, the Schwarzschild black hole cast the biggest shadow with angular diameter $\theta_d=39.6192\,\mu$as. 

\section{Conclusion}\label{sec4}
The EGB gravity theory has a number of additional nice properties than Einstein's general relativity that are not enjoyed by other higher-curvature theories.  The GB action is topological in $4D$ and hence does not contribute to gravitational dynamics in $4D$. However, using a consistent dimensional regularization procedure one can get the non-trivial contribution of the GB term to the $4D$ gravitational field equations. Furthermore, the $4D$ regularized EGB gravity theories admit spherically symmetric black holes (\ref{NR}) and depending on critical mass it has two horizons \cite{Glavan:2019inb,Cognola:2013fva}.
In this paper, we have considered the rotating black hole in the regularized $4D$ EGB gravity, which has an additional GB parameter $\alpha$ than the Kerr black hole, and it produces deviation from Kerr geometry but with a richer configuration of the event horizon and SLS. The rotating EGB black holes allow studying the effect of  higher curvature on the Kerr black holes. It is found that the GB coupling parameter $\alpha$ makes a profound influence on the structure of the horizon by reducing its radius. For a fixed value of black hole spin $a$, there always exists an extremal value of $\alpha=\alpha_E$, for which black hole has degenerate horizons i.e., $r_-=r_+=r_E$, black hole with two distinct horizons for $\alpha<\alpha_E$, and no horizon for $\alpha>\alpha_E$. Similarly, for a given value of $\alpha$, one can obtain the extremal value of $a=a_E$ for which $r_-=r_+=r_E$. The radii of horizons significantly decrease with increasing $\alpha$ and the ergosphere area is also affected, thereby can have interesting consequences on the astrophysical Penrose process.  

This persuade us to reconsider the shadow cast by the rotating EGB black holes by discussing the photons geodesics equations of motion, which are analytically solved in the first-order differential form. Observables, namely area $A$ and oblateness $D$, are used to characterize the size and shape of the shadows. It is noticed that the rotating EGB black holes cast smaller and more distorted shadows than those for the Kerr black holes. The shadow size further decreases and the distortion increases with the increasing $\alpha$. It is shown that for a given set of shadow observables, namely, area $A$ and oblateness $D$, we can explicitly determine the black hole parameters ($a,\alpha$). Shadow observational results for the M87* black hole are used to place constraints on the GB parameter in the supermassive black hole context. We modeled the M87* black hole as the rotating EGB black hole and used the inferred shadow angular diameter and the circularity deviation observables to determine the bound on $\alpha$. We have shown that within a finite parameter space, e.g. for $a=0.1M$, the rotating EGB black hole $\alpha\leq 0.00394M^2$ is consistent with the M87* shadow results.

Further, the static and spherically symmetric black hole solution of regularized $4D$ EGB gravity is identical as those found in semi-classical Einstein's equations with conformal anomaly \cite{Cai:2009ua}, gravity theory with quantum corrections \cite{Cognola:2013fva}, the regularized Lovelock gravity \cite{Casalino:2020kbt}, and the scalar-tensor Horndeski gravity theory \cite{Lu:2020iav}. Therefore, the presented study of rotating black hole is also valid for these gravity black holes.

\section{Acknowledgments} S.G.G. would like to thank DST INDO-SA bilateral project DST/INT/South Africa/P-06/2016, SERB-DST for the ASEAN project IMRC/AISTDF/CRD/2018/000042 and also IUCAA, Pune for the hospitality while this work was being done. R.K. would like to thanks UGC for providing SRF.

\textbf{Note added in proof:} After this work was completed, we learned a similar work by Wei and Liu \cite{Wei:2020ght}, which appeared in arXiv a couple of days before.

\appendix
\section{Rotating $4D$ EGB black hole}
Here, we discuss the Azreg-Ainou \cite{Azreg-Ainou:2014pra} non-complexification procedure of modified Newman-Janis algorithm for constructing stationary spacetimes from the static seed metric (\ref{NR}) 
\begin{equation}
ds^2=-f(r)dt^2+\frac{dr^2}{f(r)}+ r^2\left(d\theta^2+ \sin^2\theta d\phi^2\right).
\end{equation}
To write down the above metric in the advance null (Eddington-Finkelstein) coordinates $(u,r,\theta,\phi)$, we define the transformation
\begin{equation}
du=dt-\frac{dr}{f(r)},
\end{equation}
the static metric in the advance null coordinate becomes
\begin{equation}
ds^2=-f(r) du^2-2dudr+r^2\left(d\theta^2+\sin^2\theta d\phi^2\right).
\end{equation}
We introduce the set of null tetrad $Z_\alpha^\mu=(l^\mu,n^\mu,m^\mu,\bar{m}^\mu)$, such that the inverse metric $g^{\mu\nu}$ is defined as
\begin{equation}
g^{\mu\nu}=-l^\mu n^\nu -l^\nu n^\mu +m^\mu \bar{m}^\nu +m^\nu \bar{m}^\mu,\label{g1}
\end{equation}
with
\begin{equation}
l^\mu=\delta^\mu_r, \;\; n^\mu=\delta^\mu_u-\frac{f(r)}{2}\delta^\mu_r, \;\; m^\mu=\frac{1}{\sqrt{2}r}\left(\delta^\mu_\theta+\frac{i}{\sin\theta}\delta^\mu_\phi\right).
\end{equation}
and $\bar{m}^\mu$ is the complex conjugate of $m^\mu$. Tetrad vectors satisfy the following relations
\begin{equation}
l_\mu l^\mu = n_\mu n^\mu = m_\mu m^\mu = l_\mu m^\mu = n_\mu m^\mu =0,
\end{equation}
\begin{equation}
l_\mu n^\mu = - m_\mu \bar{m}^\mu =-1.
\end{equation}
Next we perform the complex coordinate transformation in which $\delta^{\mu}_{\nu}$ transform as \cite{Azreg-Ainou:2014pra}
\begin{equation}
\delta^{\mu}_{r}\to \delta^{\mu}_{r},\;\;\; \delta^{\mu}_{u}\to \delta^{\mu}_{u},\;\;\; \delta^{\mu}_{\theta}\to \delta^{\mu}_{r} +ia\sin\theta(\delta^{\mu}_{u}-\delta^{\mu}_{r}),\;\;\; \delta^{\mu}_{\phi}\to \delta^{\mu}_{\phi}.
\end{equation}
Here, $a$ is the black hole spin parameter. In the modified Newman-Janis algorithm \cite{Azreg-Ainou:2014pra}, the ambiguous complexification of radial coordinate is dropped, rather it is considered that the function $f(r)$ transform
to $F= F(r, a, \theta)$, and $r^2\to H(r,a, \theta)$. Following procedure given in Ref.~\cite{Azreg-Ainou:2014pra}, the transformed null tetrads read as 
\begin{equation}
l'^\mu=\delta^\mu_r, \quad n'^\mu=\delta^\mu_u-\frac{F(r,a,\theta)}{2}\delta^\mu_r,\quad
m'^\mu=\frac{1}{\sqrt{2H(r,a,\theta)}}\left(ia\sin\theta(\delta^\mu_u-\delta^\mu_r)+\delta^\mu_\theta+\frac{i}{\sin\theta}\delta^\mu_\phi\right).\label{t1}
\end{equation}
Using eq.~(\ref{t1}) in definition (\ref{g1}), the new inverse metric reads as
\begin{equation}
g^{\mu\nu}=-l'^\mu n'^\nu -l'^\nu n'^\mu +m'^\mu \bar{m}'^\nu +m'^\nu \bar{m}'^\mu,
\end{equation}
which gives the rotating black hole metric in the Eddington-Finkelstein coordinates as follow
\begin{eqnarray}\notag
ds^2&=&-F(r,a,\theta)du^2-2dudr+2a\sin^2\theta\left(F(r,a,\theta)-1\right)du d\phi+ 2a\sin^2\theta drd\phi+H(r,a,\theta) d\theta^2 \\
&&+\sin^2\theta\left[H(r,a,\theta)+a^2\sin^2\theta\left(2-F(r,a,\theta)\right)\right]d\phi^2.
\label{g3}
\end{eqnarray}
The final but important step is to bring the metric (\ref{g3}) to Boyer-Lindquist
coordinates by a global coordinate transformation. Therefore, we choose
\begin{equation}
du=dt'+\lambda(r)dr, \hspace{0.5cm} d\phi=d\phi'+\chi(r) dr,
\label{trans}
\end{equation}
such that $\lambda(r)$ and $\chi(r)$ are sole functions of $r$ only. We choose \cite{Azreg-Ainou:2014pra}
\begin{equation}
\lambda(r)=-\frac{r^2+a^2}{f(r)r^2+a^2},\;\;\; \chi(r)=-\frac{a}{f(r)r^2+a^2}.
\end{equation}
Setting the coefficient of cross-term $dtdr$ in the metric to zero, we obtain
\begin{equation}
F(r,a,\theta)=\frac{f(r)r^2+a^2 \cos^2\theta}{H(r,a,\theta)^2},
\end{equation}
and for the vanishing Einstein tensor component, $G_{r \theta}=0$, we choose $H(r,a,\cos\theta)=r^2+a^2 \cos^2\theta$. Finally, the rotating black hole metric reads as
\begin{eqnarray}
ds^2 &=&  -\left(\frac{\Delta - a^2 \sin^2 \theta}{\Sigma}\right) dt^2 + \frac{ \Sigma}{\Delta }  \, dr^2  -  2 a \sin^2 \theta \left(1 - \frac{\Delta - a^2 \sin^2 \theta}{\Sigma} \right) dt \, d \phi + \Sigma \, d \theta^2  \nonumber \\
&& +  \, \sin ^2 \theta  \left[ \Sigma + a^2 \sin^2 \theta \left(2 - \frac{\Delta - a^2 \sin^2\theta}{\Sigma}\right)   \right]    d \phi^2,
\label{rotbhr1}
\end{eqnarray}
with 
\begin{eqnarray}
\Delta=a^2+r^2f(r),\;\;\; \Sigma=r^2+a^2 \cos^2\theta.
\end{eqnarray}

\noindent

\begin{thebibliography}{99}

\bibitem{Lovelock:1972vz}
D.~Lovelock,
J.\ Math.\ Phys.\  {\bf 13} 874 (1972).


\bibitem{Lanczos:1938sf}
C.~Lanczos,
Annals Math.\  {\bf 39} 842 (1938).


\bibitem{Lovelock:1971yv}
D.~Lovelock,
J.\ Math.\ Phys.\  {\bf 12}  498 (1971).


\bibitem{Zwiebach:1985uq} 
B.~Zwiebach,
Phys.\ Lett.\  {\bf 156B}, 315 (1985).

\bibitem{Nojiri:2018ouv} 
S.~Nojiri, S.~D.~Odintsov and V.~K.~Oikonomou,
Phys.\ Rev.\ D {\bf 99}, 044050 (2019).


\bibitem{Wiltshire:1985us} 
D.~L.~Wiltshire,
Phys.\ Lett.\ B {\bf 169}, 36 (1986).


\bibitem{Boulware:1985wk}  D.G.~Boulware and S.~Deser, Phys. Rev. Lett. {\bf 55}, 2656 (1985); 
J.T.~Wheeler, Nucl. Phys. B {\bf 268}, 737 (1986). 

\bibitem {egb}
S.~Nojiri and S.~D.~Odintsov,
Phys.\ Lett.\ B {\bf 521} 87 (2001);
Erratum: [Phys.\ Lett.\ B {\bf 542} 301 (2002)];
Y. M. Cho and I. P. Neupane, Phys. Rev. D {\bf 66}, 024044 (2002); 
M. Cvetic, S. Nojiri and S. D. Odintsov, Nucl. Phys. {\bf B 628}, 295 (2002); 
R. G. Cai, Phys. Rev. D {\bf 65}, 084014 (2002); 
I. P. Neupane, Phys. Rev. D {\bf 67}, 061501(R) (2003); {\bf 69}, 084011 (2004);  
A. Padilla, Classical Quantum Gravity {\bf 20}, 3129 (2003); 
N. Deruelle, J. Katz, and S. Ogushi, Classical Quantum Gravity {\bf 21}, 1971 (2004); 
M. H. Dehghani, Phys. Rev. D {\bf 69}, 064024 (2004);  
R. G. Cai and Q. Guo, Phys. Rev. D {\bf 69}, 104025 (2004); 
T. Torii and H. Maeda, Phys. Rev. D {\bf 71}, 124002 (2005); 
M. H. Dehghani and R. B. Mann, Phys. Rev. D {\bf 72}, 124006 (2005); 
M. H. Dehghani and S. H. Hendi, Phys. Rev. D {\bf 73}, 084021 (2006); 
M. H. Dehghani, G. H. Bordbar, and M. Shamirzaie, Phys. Rev. D {\bf 74}, 064023 (2006).



\bibitem{ghosh} S.~Jhingan and S.~G.~Ghosh,
Phys.\ Rev.\ D {\bf 81}, 024010 (2010);
S.~G.~Ghosh, M.~Amir and S.~D.~Maharaj,
Eur.\ Phys.\ J.\ C {\bf 77}, 530 (2017);
S.~G.~Ghosh,
Class.\ Quant.\ Grav.\  {\bf 35}, 085008 (2018). 
\bibitem{egb2} 
S.~Mignemi and N.~R.~Stewart,
Phys.\ Rev.\ D {\bf 47}, 5259 (1993);
P.~Kanti, N.~E.~Mavromatos, J.~Rizos, K.~Tamvakis and E.~Winstanley,
Phys.\ Rev.\ D {\bf 54}, 5049 (1996);
S.~O.~Alexeev and M.~V.~Pomazanov,
Phys.\ Rev.\ D {\bf 55}, 2110 (1997);
T.~Torii, H.~Yajima and K.~i.~Maeda,
Phys.\ Rev.\ D {\bf 55}, 739 (1997);
R.~Konoplya,
Phys.\ Rev.\ D {\bf 71}, 024038 (2005);
B.~Kleihaus, J.~Kunz and E.~Radu,
Phys.\ Rev.\ Lett.\  {\bf 106}, 151104 (2011);
A.~Maselli, P.~Pani, L.~Gualtieri and V.~Ferrari,
Phys.\ Rev.\ D {\bf 92}, 083014 (2015).


\bibitem{Ghosh1:2018bxg} 
S.~G.~Ghosh, D.~V.~Singh and S.~D.~Maharaj,
Phys.\ Rev.\ D {\bf 97}, 104050 (2018);
S.~Hyun and C.~H.~Nam,
Eur.\ Phys.\ J.\ C {\bf 79},737 (2019);
A.~Kumar, D.~Veer Singh and S.~G.~Ghosh,
Eur.\ Phys.\ J.\ C {\bf 79}, 275 (2019);
D.~V.~Singh, S.~G.~Ghosh and S.~D.~Maharaj,
Annals Phys.\  {\bf 412}, 168025 (2020).

\bibitem{Tomozawa:2011gp}
Y.~Tomozawa,
arXiv:1107.1424 [gr-qc].


\bibitem{Cognola:2013fva} 
G.~Cognola, R.~Myrzakulov, L.~Sebastiani and S.~Zerbini,
Phys.\ Rev.\ D {\bf 88}, 024006 (2013).


\bibitem{Glavan:2019inb}
D.~Glavan and C.~Lin,
Phys.\ Rev.\ Lett.\  {\bf 124}, 081301 (2020).


\bibitem{Guo:2020zmf} 
M.~Guo and P.~C.~Li,
arXiv:2003.02523 [gr-qc].


\bibitem{Wei:2020ght} 
S.~W.~Wei and Y.~X.~Liu,
arXiv:2003.07769 [gr-qc].


\bibitem{Konoplya:2020bxa}
R.~A.~Konoplya and A.~F.~Zinhailo,
arXiv:2003.01188 [gr-qc].


\bibitem{Fernandes:2020rpa} 
P.~G.~S.~Fernandes,
arXiv:2003.05491 [gr-qc].


\bibitem{Doneva:2020ped} 
D.~D.~Doneva and S.~S.~Yazadjiev,
arXiv:2003.10284 [gr-qc].


\bibitem{Ghosh:2020vpc} 
S.~G.~Ghosh and S.~D.~Maharaj,
Physics of the Dark Universe 30, 100687 (2020).

\bibitem{Singh:2020mty}
D.~V.~Singh, R.~Kumar, S.~G.~Ghosh and S.~D.~Maharaj,
arXiv:2006.00594 [gr-qc].

\bibitem{Kumar:2020sag}
R.~Kumar, S.~U.~Islam and S.~G.~Ghosh,
arXiv:2004.12970 [gr-qc].

\bibitem{Islam:2020xmy}
S.~U.~Islam, R.~Kumar and S.~G.~Ghosh,
arXiv:2004.01038 [gr-qc].

\bibitem{Ghosh:2020syx}
S.~G.~Ghosh and R.~Kumar,
arXiv:2003.12291 [gr-qc].

\bibitem{Konoplya:2020qqh}
R.~A.~Konoplya and A.~Zhidenko,
Phys. Rev. D \textbf{101}, 084038 (2020).


\bibitem{Casalino:2020kbt} 
A.~Casalino, A.~Colleaux, M.~Rinaldi and S.~Vicentini,
arXiv:2003.07068 [gr-qc].

\bibitem{Hennigar:2020lsl}
R.~A.~Hennigar, D.~Kubiznak, R.~B.~Mann and C.~Pollack,
arXiv:2004.09472 [gr-qc].

\bibitem{Mahapatra:2020rds}
S.~Mahapatra,
[arXiv:2004.09214 [gr-qc]].

\bibitem{Ai:2020peo}
W.~Ai,
arXiv:2004.02858 [gr-qc].


\bibitem{Shu:2020cjw}
F.~Shu,
arXiv:2004.09339 [gr-qc].


\bibitem{Gurses:2020ofy}
M.~Gurses, T.~C.~Sisman and B.~Tekin,
arXiv:2004.03390 [gr-qc].

\bibitem{Fernandes:2020nbq}
P.~G.~Fernandes, P.~Carrilho, T.~Clifton and D.~J.~Mulryne,
arXiv:2004.08362 [gr-qc].

\bibitem{Lu:2020iav}
H.~Lu and Y.~Pang,
arXiv:2003.11552 [gr-qc].


\bibitem{Kobayashi:2020wqy}
T.~Kobayashi,
arXiv:2003.12771 [gr-qc].


\bibitem{Mann}
R. B. Mann and S. F. Ross, 
Class. Quant. Grav. 10 (1993) 1405.

\bibitem{Cai:2009ua} 
R.~G.~Cai, L.~M.~Cao and N.~Ohta,
JHEP {\bf 1004}, 082 (2010);
R.~G.~Cai,
Phys.\ Lett.\ B {\bf 733}, 183 (2014).


\bibitem{Kehagias:2009is}
A.~Kehagias and K.~Sfetsos,
Phys. Lett. B \textbf{678}, 123 (2009).

\bibitem{Newman:1965tw}
E.~Newman and A.~Janis,
J.\ Math.\ Phys.\  \textbf{6}, 915 (1965).


\bibitem{Johannsen:2011dh}
T.~Johannsen and D.~Psaltis,
Phys.\ Rev.\ D \textbf{83}, 124015 (2011).

\bibitem{Jusufi:2019caq}
K.~Jusufi, M.~Jamil, H.~Chakrabarty, Q.~Wu, C.~Bambi and A.~Wang,
Phys.\ Rev.\ D \textbf{101}, 044035 (2020).

\bibitem{Bambi:2013ufa}
C.~Bambi and L.~Modesto,
Phys.\ Lett.\ B \textbf{721}, 329 (2013).


\bibitem{Moffat:2014aja} 
J.~W.~Moffat,
Eur.\ Phys.\ J.\ C {\bf 75}, 175 (2015).

\bibitem{Ghosh:2014hea} 
S.~G.~Ghosh and S.~D.~Maharaj,
Eur.\ Phys.\ J.\ C {\bf 75}, 7 (2015).

\bibitem{Hansen:2013owa} 
D.~Hansen and N.~Yunes,
Phys.\ Rev.\ D {\bf 88}, 104020 (2013).

\bibitem{Azreg-Ainou:2014pra} 
M.~Azreg-A{\"i}nou,
Phys.\ Rev.\ D {\bf 90}, 064041 (2014).

\bibitem{Azreg-Ainou:2014aqa} 
M.~Azreg-A{\"i}nou,
Eur.\ Phys.\ J.\ C {\bf 74}, 2865 (2014).



\bibitem{Kerr:1963ud} 
R.~P.~Kerr,
Phys.\ Rev.\ Lett.\  {\bf 11}, 237 (1963).


\bibitem {Chandrasekhar:1992}
S.~ Chandrasekhar, 
{\it The Mathematical Theory of Black Holes} (Oxford University Press, New York, 1992).


\bibitem{pen} R. Penrose, Riv. Nuovo Cimento Numero Speciale {\bf 1}, 252 (1969)[Gen. Relativ. Gravit. {\bf 34}, 1141 (2002)].

\bibitem {Frolov:1998wf}
V.~P.~Frolov and A.~Zelnikov, 
{\it Introduction to Black Hole Physics} (Oxford University Press, New York, 2011).

\bibitem{Synge:1966} J.~L.~Synge,
Mon.\ Not.\ R.\ Astron.\ Soc.\  {\bf 131}, 463 (1966).

\bibitem{Luminet:1979} J. P. Luminet, Astron. Astrophys. {\bf 75}, 228 (1979).

\bibitem{Bardeen} J. M. Bardeen, \textit{ Black Holes}, Edited
by C. DeWitt and B. S. DeWitt (Gordon and Breach,
New York, 1973, p. 215).


\bibitem{CT} C. T. Cunningham, J. M. Bardeen, Astrophys. J. {\bf 173}, L137 (1972).

\bibitem{De1} A. de Vries,  Class.\ Quant.\ Grav.\ {\bf 17}, 123 (2000); 
Z. Q. Shen, K.Y. Lo, M. C. Liang, P. T. P. Ho and J. H. Zhao, Nature \textbf{438}, 62 (2005);
L.~Amarilla, E.~F.~Eiroa and G.~Giribet,
Phys.\ Rev.\ D {\bf 81}, 124045 (2010);
L.~Amarilla and E.~F.~Eiroa,
Phys.\ Rev.\ D {\bf 85}, 064019 (2012);
A.~Yumoto, D.~Nitta, T.~Chiba and N.~Sugiyama,
Phys.\ Rev.\ D {\bf 86}, 103001 (2012);
L.~Amarilla and E.~F.~Eiroa,
Phys.\ Rev.\ D {\bf 87}, 044057 (2013);
F.~Atamurotov, A.~Abdujabbarov, and B.~Ahmedov,
Phys.\ Rev.\ D {\bf 88}, 064004 (2013);
A.~Abdujabbarov, M.~Amir, B.~Ahmedov and S.~G.~Ghosh,
Phys.\ Rev.\ D {\bf 93}, 104004 (2016);
M.~Amir, B.~P.~Singh and S.~G.~Ghosh,
Eur.\ Phys.\ J.\ C {\bf 78}, 399 (2018).




\bibitem{Grenzebach1} 
A.~A.~Abdujabbarov, L.~Rezzolla, and B.~J.~Ahmedov,
Mon.\ Not.\ R.\ Astron.\ Soc.\  {\bf 454}, 2423 (2015);
P.~V.~P.~Cunha and C.~A.~R.~Herdeiro,
Gen.\ Rel.\ Grav.\  {\bf 50}, 42 (2018);
Y.~Mizuno {\it et al.},
Nat.\ Astron.\  {\bf 2}, 585 (2018);
R.~Shaikh,
Phys.\ Rev.\ D {\bf 100}, 024028 (2019);
A.~K.~Mishra, S.~Chakraborty and S.~Sarkar,
Phys.\ Rev.\ D {\bf 99}, 104080 (2019);
R.~Kumar, A. Kumar, and S.~G.~Ghosh,
Astrophys.\ J.\  {\bf 896}, 89 (2020).

\bibitem{Kumar:2018ple} 
R.~Kumar and S.~G.~Ghosh,
Astrophys.\ J.\  {\bf 892}, 78 (2020).


\bibitem{Kramer1:2004hd} 
M.~Kramer, D.~C.~Backer, J.~M.~Cordes, T.~J.~W.~Lazio, B.~W.~Stappers and S.~Johnston,
New Astron.\ Rev.\  {\bf 48}, 993 (2004);
D.~Psaltis,
Living Rev.\ Rel.\  {\bf 11}, 9 (2008);
T.~Harko, Z.~Kovacs and F.~S.~N.~Lobo,
Phys.\ Rev.\ D {\bf 80}, 044021 (2009);
D.~Psaltis, F.~Ozel, C.~K.~Chan and D.~P.~Marrone,
Astrophys.\ J.\  {\bf 814}, 115 (2015);
T.~Johannsen {\it et al.},
Phys.\ Rev.\ Lett.\  {\bf 116}, 031101 (2016);
D.~Psaltis,
Gen.\ Rel.\ Grav.\  {\bf 51}, 137 (2019);




\bibitem{Carter:1968rr} 
B.~Carter, 
Phys.\ Rev.\ D {\bf 174}, 1559 (1968).  

\bibitem{Grenzebach} A.~Grenzebach, V.~Perlick, and C.~L{\"a}mmerzahl,
Phys.\ Rev.\ D {\bf 89}, 124004 (2014).



\bibitem{Tsupko:2017rdo} 
O.~Y.~Tsupko,
Phys.\ Rev.\ D {\bf 95}, 104058 (2017).

\bibitem{Johannsen:2010ru} 
T.~Johannsen and D.~Psaltis,
Astrophys.\ J.\  {\bf 718}, 446 (2010).

\bibitem{Johannsen:2015qca} 
T.~Johannsen,
Astrophys.\ J.\  {\bf 777}, 170 (2013).


\bibitem{EHT}
https://eventhorizontelescope.org/


\bibitem{Akiyama:2019cqa} 
K.~Akiyama {\it et al.},
Astrophys.\ J.\  {\bf 875}, L1 (2019).


\bibitem{Akiyama:2019eap} 
K.~Akiyama {\it et al.},
Astrophys.\ J.\  {\bf 875}, L6 (2019).


\bibitem{Bambi1} 
C.~Bambi, K.~Freese, S.~Vagnozzi and L.~Visinelli,
Phys.\ Rev.\ D {\bf 100}, 044057 (2019);
S.~Vagnozzi and L.~Visinelli,
Phys.\ Rev.\ D {\bf 100}, 024020 (2019);
R.~Kumar, S.~G.~Ghosh and A.~Wang,
Phys.\ Rev.\ D {\bf 100}, 124024 (2019);
P.~V.~P.~Cunha, C.~A.~R.~Herdeiro and E.~Radu,
Universe {\bf 5}, 220 (2019);
R.~Kumar, B.~P.~Singh and S.~G.~Ghosh,
arXiv:1904.07652 [gr-qc];
I.~Banerjee, S.~Chakraborty and S.~SenGupta,
arXiv:1909.09385 [gr-qc];
I.~Banerjee, S.~Sau and S.~SenGupta,
arXiv:1911.05385 [gr-qc];
A.~Allahyari, M.~Khodadi, S.~Vagnozzi and D.~F.~Mota,
arXiv:1912.08231 [gr-qc].
R.~Kumar, S.~G.~Ghosh and A.~Wang,
Phys. Rev. D \textbf{101}, 104001 (2020).
A.~Narang, S.~Mohanty and A.~Kumar,
arXiv:2002.12786 [gr-qc].
J.~W.~Moffat and V.~T.~Toth,
Phys.\ Rev.\ D {\bf 101}, 024014 (2020).
S.~G.~Ghosh,
Eur. Phys. J. C \textbf{75}, 532 (2015);
N.~Dadhich and S.~G.~Ghosh,
arXiv:1307.6166 [gr-qc];
S.~G.~Ghosh, S.~D.~Maharaj and U.~Papnoi,
Eur. Phys. J. C \textbf{73}, 2473 (2013).




\end{thebibliography}
\end{document}